\documentclass[12pt, preprint]{aastex}
\usepackage{graphicx}
\def\max{{\rm max}}
\def\min{{\rm min}}

\def\eff{{\rm eff}}

\def\e{{\rm E}}

\begin{document}
\title{Korea Microlensing Telescope Network Microlensing Events from 2015: Event-Finding Algorithm, Vetting, and Photometry}

\author{\textsc{D.-J. Kim$^{1}$, H.-W. Kim$^{1}$, K.-H. Hwang$^{1}$ \\
and \\ M. D. Albrow$^{2}$, S.-J.Chung$^{1}$, A. Gould$^{1,3,4}$, C. Han$^{5}$, Y.
K. Jung$^{6}$, Y.-H. Ryu$^{1}$, I.-G. Shin$^{6}$, J. C. Yee$^{6}$, W. Zhu$^{4}$, S.-M. Cha$^{1}$, S.-L. Kim$^{1}$, C.-U. Lee$^{1}$, D.-J. Lee$^{1}$, Y. Lee$^{1}$, B.-G. Park$^{1}$, R. W. Pogge$^{4}$ \\ (The KMTNet Collaboration)\\ }}

\affil{$^{1}$ Korea Astronomy and Space Science Institute, Daejeon 34055, Republic of Korea \\
$^{2}$ University of Canterbury, Department of Physics and Astronomy, Private Bag 3800, Christchurch 8020, New Zealand \\
$^{3}$ Max-Planck-Institute for Astronomy, K\"{o}nigstuhl 16, 59116 Heidelberg, Germany \\
$^{4}$ Department of Astronomy, Ohio State University, 130 W. 18th Ave., Columbus, OH 32210, USA \\
$^{5}$ Department of Physics, Chungbuk National University, Cheongju 28644, Republic of Korea \\
$^{6}$ Harvard-Smithsonian Center for Astrophysics, 50 Garden St., Cambridge, MA 02128, USA}

\begin{abstract}
We present microlensing events in the 2015 Korea Microlensing
Telescope Network (KMTNet) data and our procedure for identifying these events.
In particular, candidates were detected with a novel ``completed
event'' microlensing event-finder algorithm.  The algorithm works
by making linear fits to a $(t_0,t_\eff,u_0)$ grid of point-lens
microlensing models.  This approach is rendered computationally
efficient by restricting $u_0$ to just two values (0 and 1), which we
show is quite adequate.  The implementation presented here is
specifically tailored to the commission-year character of the 2015
data, but the algorithm is quite general and has already been applied
to a completely different (non-KMTNet) data set.  We outline expected
improvements for 2016 and future KMTNet data.  The light curves of the
660 ``clear microlensing'' and 182 ``possible microlensing'' events
that were found in 2015 are presented along with our policy for
their public release.
\end{abstract}


\section{{Introduction}
\label{sec:intro}}

\citet{gouldloeb} originally advocated a two-step approach for finding
planets by gravitational microlensing.  In the first step, a wide area
survey would monitor 10s or 100s of million of stars roughly once per
night using a wide-angle camera in order to find microlensing events.
Then, in the second step, individual events found in the first step
would be monitored at high cadence from several longitudes using
narrow angle cameras.  That is, the cadence of each survey would be
matched to the timescale of the effects being sought: a few dozen
points during the roughly month-long microlensing events from the
first survey and a few dozen points over the day-long (or shorter)
planetary perturbations from the second survey.  As \citet{gouldloeb}
specifically pointed out, this approach required ``microlensing
alerts'', i.e., the recognition and public notification of
microlensing events while they were still in progress, and preferably
before they peaked.

This approach was in fact adopted by the microlensing community and
led to the detection of several dozen planets.  Right from the
beginning, however, (including the very first microlensing planet
OGLE-2003-BLG-235, \citealt{ob03235}) planets were detected by the
surveys themselves, without any ``second step'' follow-up
observations.  As the surveys went through several generations of
improvements, such survey-only detections became more common, e.g.,
\citet{ob120406}.  Nevertheless, given that many of the planets found
do require a second step, the same ``microlensing alert'' mode of
event detection, pioneered by the Optical Gravitational Lensing
Experiment (OGLE) group \citep{ews1,ews2}, remained the main practical
method by which events were discovered.

Although the ``microlensing alert'' system is the main path to event
detection, the alternate approach of finding ``completed events'' in
the data set has been present from the birth of the field.  Both the
MACHO and EROS collaborations developed algorithms for searching
through archival data to find microlensing events \citep{alcock97, eros2, eros2_2}, and in particular, the very first microlensing
event MACHO-LMC-1 was found this way \citep{lmc1}.  One important
advantage of this approach is that it has relatively little reliance
on human input and therefore can (mostly) be modeled as an objective
algorithm, which then permits objective estimates of microlensing
event rates (or planet rates). Such algorithms have been used to
measure the microlensing optical depth in the LMC and SMC
\citep{wyr09, wyr10, wyr11a, wyr11b}. \citet{wyr15} specifically
applied this approach to OGLE-III observations of the Galactic bulge,
which is the main (so far, only) field where microlensing planets are
discovered.

Directly opposite the \citet{gouldloeb} observing regime would be a
very high-cadence survey with multiple sites allowing continuous
monitoring of the Galactic bulge without the need for any followup
observations. The first survey of this type was a collaboration
between the OGLE, MOA, and Wise observatories
\citep{shvartzvald12,shvartzvald14}.

As originally conceived, the Korea Microlensing Telescope Network
(KMTNet, \citealt{kmtnet}) would also lie in this regime. KMTNet
consists of three 1.6m telescopes, each equipped with $4\,{\rm deg}^2$
cameras, and located on three southern continents, CTIO (KMTC, South
America), SAAO (KMTS, Africa), and SSO (KMTA, Australia).  According
to the original plan, which was basically implemented in 2015, it
would observe four fields ($16\,{\rm deg}^2$) continuously with a
cadence of $\Gamma=6\,{\rm hr}^{-1}$.  Hence, there would be virtually
no point in follow-up observations, and therefore no point in
microlensing alerts.  This in turn implied that KMTNet should focus on
finding completed events, both because it is easier than finding
events in real time and because (as noted above) of the potential of such an
event-finding algorithm for measuring rates.

In fact, the above paragraph notwithstanding, there would be many
potential applications for KMTNet alerts.  Most of these stem from the
fact that KMTNet has abandoned its original strategy as implemented in
2015, in favor of the layered approach pioneered by the OGLE group.
Currently, (3,7,11,3) fields are observed at cadences
$\Gamma=(4,1,0.4,0.2)\,{\rm hr}^{-1}$.  In all but the highest cadence
fields, high-magnification events could be profitably followed-up at
substantially higher cadence.  Moreover, if anomalies could be alerted
in real time in these fields, then detected planets could be
characterized much better.  Actually, alerts for the highest cadences
fields would be exceptionally important for the next few years due to
the emergence of {\it Spitzer} microlensing, which was not at all
anticipated at the time KMTNet was conceived in 2004.  Because {\it
  Spitzer} is in solar orbit, synoptic {\it Spitzer} observations can
yield ``microlens parallaxes'', which are critical for characterizing
the microlens and any planets it may have
\citep{ob140124,yee15,prop2013,prop2014,prop2015a,prop2015b,prop2016}.
However, in order for {\it Spitzer} to take useful observations, it
must be alerted to the microlens target before the event ends, and
usually before peak.  Hence, in sharp contrast to the situation
envisaged a decade ago, both a ``completed-event'' finder and alert
capability are important for KMTNet.

Ideally, therefore, both forms of event-finder would have
been basically ready when the first KMTNet data began arriving from
the telescope in early 2015, or certainly by early 2016 following the
first year of commissioning data.  Unfortunately, work on
event-finders did not begin until mid-2016. By the time the event
finder was first tested, it was late 2016, meaning that two full years
of data were already taken.  This fact alone implied that much higher
priority had to be given to constructing an event finder that worked
on completed events than to developing alert capability.  Combined
with the facts that the original core of KMTNet science was built
around a pure-survey detection strategy, and that fine-tuning and
operating a completed-event finder is much easier and less
time-consuming than an alert system, this made the completed-event
finder the obvious first choice for development.

In this paper, we present the first microlensing events detected
directly from the KMTNet survey data. In Sections \ref{sec:algo} and
\ref{sec:robust}, we present the basic algorithm for identifying
microlensing event candidates and discuss its robustness. In Section
\ref{sec:2015data}, we detail the specific procedure for applying this
algorithm to the 2015 KMTNet data starting with the photometric
pipeline and proceeding through the application of this algorithm and
the vetting of the event candidates. We also assess the efficacy of
the event-finding process by comparison to OGLE-IV EWS alerts and by
checking the detectability of binaries with our algorithm. A summary
of the detected events is given in Section \ref{sec:esum}. Information
on accessing the final sample of events is given in Section
\ref{sec:policy} along with our data policy. We discuss some potential
improvements to the algorithm in Section \ref{sec:improve} and
summarize our findings in Section \ref{sec:discuss}.

\section{The Basic Algorithm}
\label{sec:algo}

The main challenge of an event-finder is that it must be both quick
and robust.  ``Quick'' here has two senses.  First, it must sift
though $10^{12}$ photometric observations each year, spread over
$3\times 10^8$ light curves, in a reasonable amount of computer time.
Second, it must show a restricted subsample to an operator who can vet
these candidates in a reasonable amount of time.  And it must find all
candidates that might plausibly harbor a planetary signal.

By far, the fastest method of evaluating light curves is a linear fit,
which applies a simple deterministic formula twice: once to determine
the parameters and the second time to evaluate $\chi^2$.
Unfortunately, microlensing events are described by five parameters,
only two of which are linear:
\begin{equation}
F(t) = f_s A[u(t;t_0,u_0,t_\e)] + f_b;
\qquad
u(t) = \sqrt{{(t-t_0)^2\over t_\e^2} + u_0^2};
\qquad
A(u) = {u^2 + 2\over u\sqrt{u^2+4}}.
\label{eqn:foft}
\end{equation}
Here $F$ is the observed flux, $A$ is the magnification, $t_0$ is the
time of maximum magnification, $u_0$ is the impact parameter
(normalized to the Einstein radius $\theta_\e$), $t_\e$ is the
Einstein timescale, $f_s$ is the source flux, and $f_b$ is any blended
flux that does not participate in the event.

Hence, if one wishes to use a linear fit, one must work on a
3-dimensional (3D) grid of $(t_0,u_0,t_\e)$, which is prohibitive.  We
therefore begin with the insight of \citet{gould96} that in the
high-magnification limit, microlensing events can be described by just
two non-linear parameters
\begin{equation}
F(t) = {f_\max \over \sqrt{1 + (t - t_0)^2/t_\eff^2}} + f_b
\label{eqn:highmag}
\end{equation}
where $t_\eff \rightarrow u_0 t_\e$ and $f_\max \rightarrow f_s/u_0$.
This formula is approximately valid only for $u\la 0.5$, and so cannot
be applied to the entire light curve, but only to a segment around
peak.  \citet{gould96} therefore advocated a search for high
magnification events based on a 2D grid and
Equation~(\ref{eqn:highmag}), with each trial restricted to a few
effective timescales $t_\eff$ around its peak time $t_0$.

\citet{gould96} was only concerned with high-magnification events
because that was all he thought could be detected in M31, the subject
of his paper.  Whether or not this is true of M31, it is certainly not
true of the Galactic bulge, where we expect to detect many low
magnification events as well.  Therefore, we augment the
\citet{gould96} approach by considering also a representative low
magnification event, namely Equation~(\ref{eqn:foft}) with $u_0=1$.
That is, we consider
\begin{equation}
F(t) = f_1 A_j[Q(t;t_0,t_\eff)] + f_0;
\quad Q(t;t_0,t_\eff)\equiv 1 + \biggl({t-t_0\over t_\eff}\biggr)^2;
\quad (j=1,2)
\label{eqn:general}
\end{equation}
where
\begin{equation}
A_{j=1}(Q) = Q^{-1/2};
\qquad
A_{j=2}(Q) = {Q + 2\over \sqrt{Q(Q+4)}} = [1 - (Q/2 +1)^{-2}]^{-1/2}
\label{eqn:ajoftau}
\end{equation}

Note that we no longer call the two flux parameters $(f_s,f_b)$, but
rather $(f_0,f_1)$.  This is because no physical meaning can be
ascribed to these parameters.  In the high-magnification limit,
$f_1\rightarrow f_\max$ and at $u_0=1$, $(f_1,f_0)\rightarrow
(f_s,f_b)$, but in the general case, these flux parameters are not
identifiable with any specific physical quantities.

The set of $t_{\eff,k}$ are a geometric series
\begin{equation}
t_{\eff,k+1} = (1+\delta_{t_\eff})t_{\eff,k}
\label{eqn:teffseries}
\end{equation}
and for each $t_{\eff,k}$, we choose an arithmetic series for $t_{0,k,l}$
\begin{equation}
t_{0,k,l+1} = t_{0,k,l} +\delta_{t_0}t_{\eff,k}.
\label{eqn:t0series}
\end{equation}
We adopt
\begin{equation}
\delta_{t_0} = \delta_{t_{\eff}} = 1/3,
\label{eqn:deltavals}
\end{equation}
which we show below to be quite conservative.  We set
$t_{\eff,3}=1\,$day (so $t_{\eff,1}\simeq 0.56\,$day), with a maximum
$t_\eff\simeq 99\,$days.  It is not possible to reliably detect events
that are substantially longer than this upper limit in a single
season.  The lower limit is also conditioned by the characteristics of
the first season's data.  We discuss extending these limits in
Section~\ref{sec:improve}.  We consider values of $t_0$ from
$\delta_{t_0}$ before the first epoch of the 2015 season until
$\delta_{t_0}$ after the last epoch.

We restrict the fits to data within $t_0\pm Z t_\eff$ where $Z=5$.  We
require that this interval contain at least $N_\min= 50$ points.  For
2015 data, we consider only data taken at CTIO in the initial,
automated phase of the search.  We discuss expanding the search to
include other observatories in Section~\ref{sec:improve}.

There are therefore about $2\Delta T/(\delta_{t_0}t_{\eff,k})$ trial
fits for each value of $t_{\eff,k}$ and these trials will contain a
total of $4ZN/\delta_{t_0} = 240,000$ data points (with
repeats).  Here $\Delta T\sim 250\,$days is the duration of the 2015
season and $N\sim 4000$ is the total number of data points taken.
Since the computation time is basically proportional to the number of
data points, this means that each $t_\eff$ trial takes about the same
amount of time.  The total number of trials is roughly
$2(1+\delta_{t_\eff}^{-1})\Delta T/(\delta_{t_0}t_{\eff,\min})\sim 10,000$.

\section{Analytic Characterization of Robustness}
\label{sec:robust}

Since we are sampling a continuous function of three variables
$(t_0,t_\eff,u_0)$ on a discrete grid, the $\Delta\chi^2$ of the fit
(relative to a flat light-curve model) will inevitably be smaller than
for the optimal values of $(t_0, t_\eff, u_0)$. This is not in itself
of any concern because the only purpose of these fits is to find which
of the light curves have sufficient deviation to warrant showing them
to the operator.  If, for example, the discrete-model $\Delta\chi^2$
could be as much as 10\% smaller than the true (continuous model)
$\Delta\chi^2$, then one simply has to set the $\Delta\chi^2$
threshold 10\% lower than the level that one wishes to investigate.
Here we show how the discretization of the parameter space impacts the
fits.

\begin{figure}
\plotone{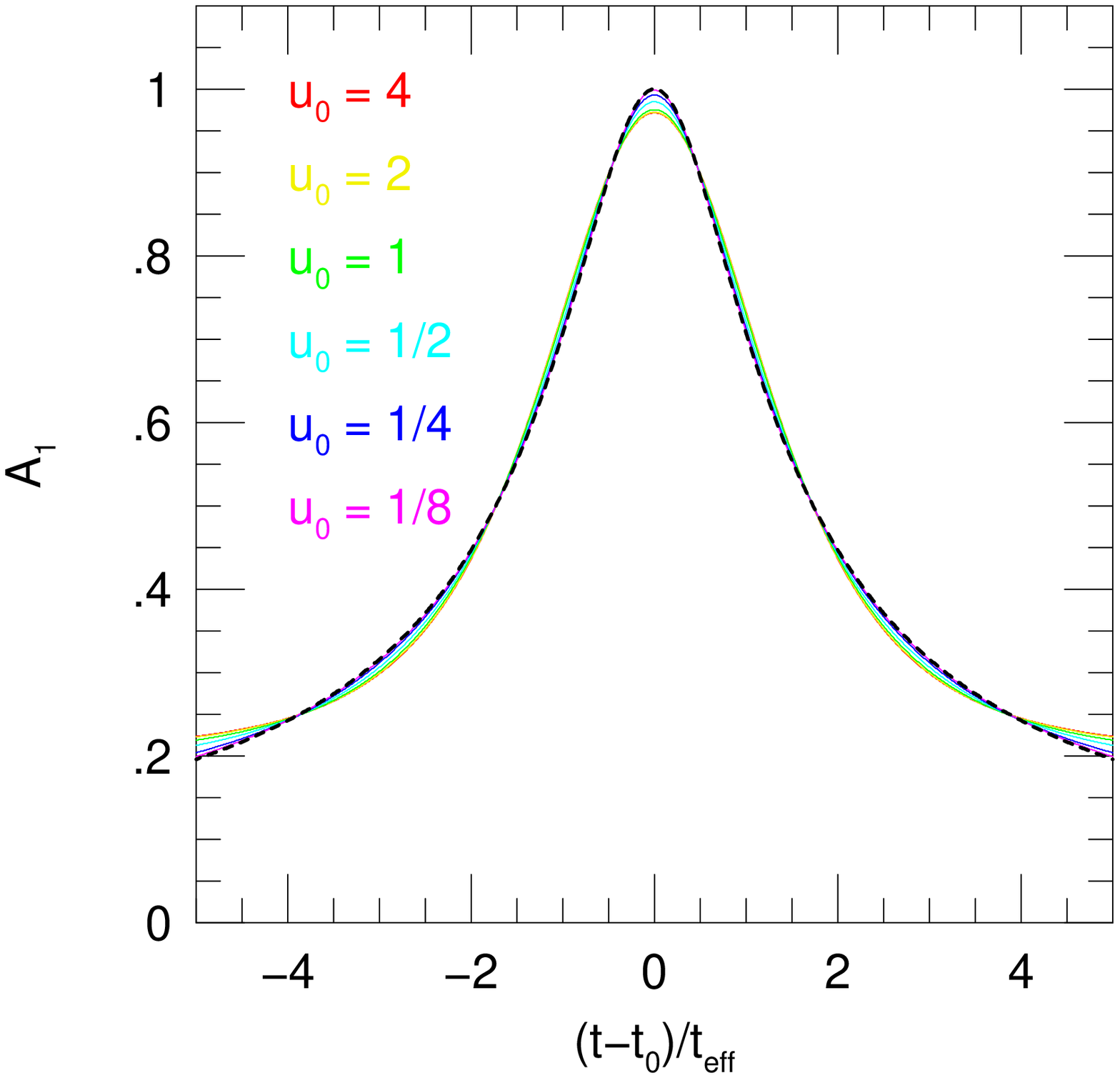}
\caption{Best-fit ``high-magnification'' $(u_0\ll 1)$ models for
various values of $u_0=4,2,1,1/2,1/4,1/8$ from among all possible
values of $(t_0,t_\eff)$.  Surprisingly, even very low mag models
fit reasonably well.  Moreover $u_0\leq 1/4$ fit extremely well,
and even $u_0= 1/2$ fits quite well. Dashed black line shows
the model functional form.
}
\label{fig:hm}
\end{figure}

\begin{figure}
\plotone{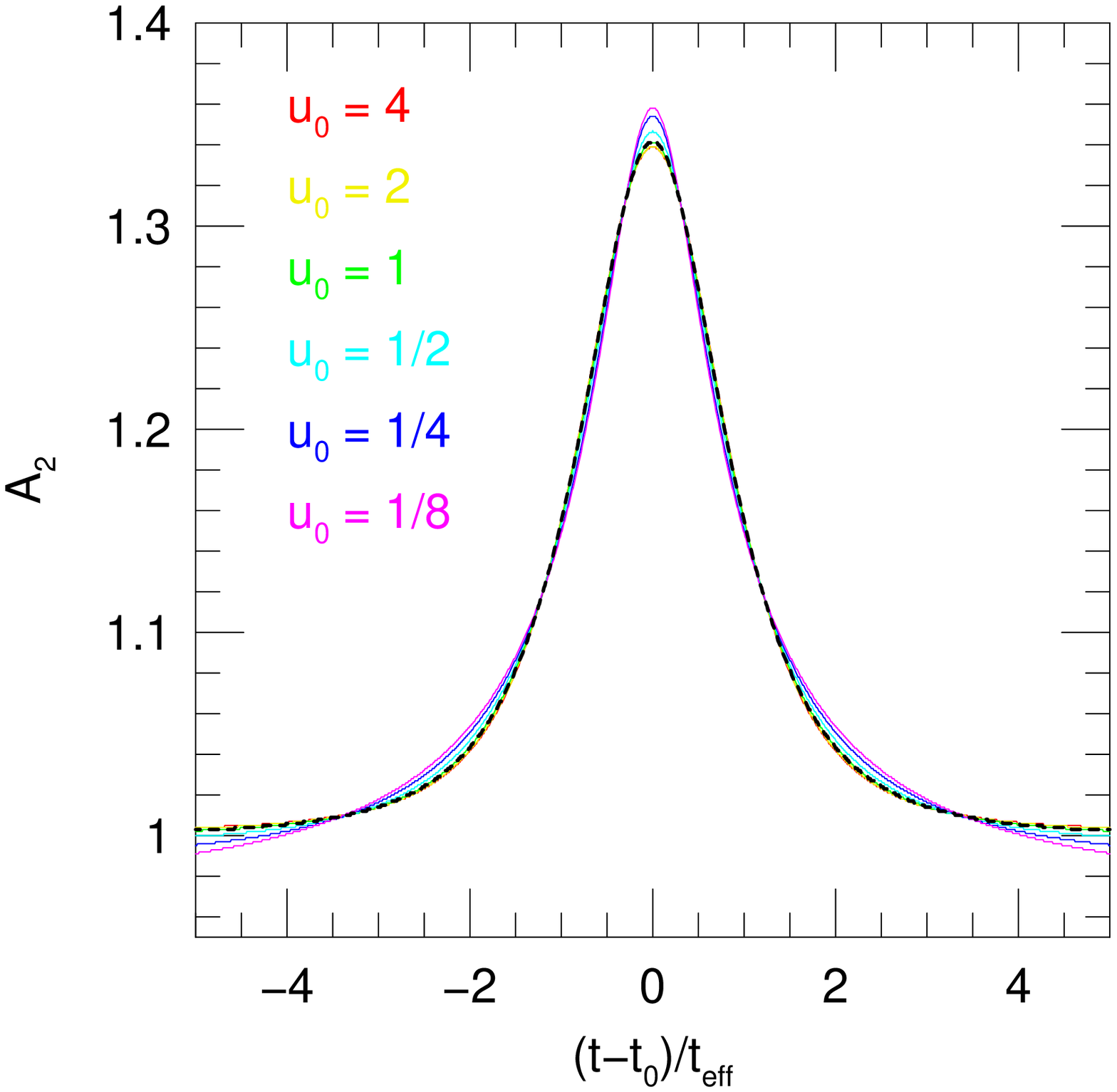}
\caption{Best-fit ``low-magnification'' $(u_0= 1)$ models for
various values of $u_0=4,2,1,1/2,1/4,1/8$ from among all possible
values of $(t_0,t_\eff)$.  Surprisingly, even very high mag models
fit reasonably well.  Moreover $u_0\geq 1$ fit extremely well,
and even $u_0= 1/2$ fits quite well. Dashed black line shows
the model functional form.
}
\label{fig:lm}
\end{figure}

We would expect the largest $\Delta\chi^2$ deviation from the most
coarsely sampled parameter $u_0$. As outlined in
Section~\ref{sec:algo}, $u_0$ is represented by just two discrete
values: ``high-mag'' ($u_0\ll 1$, in effect, ``$u_0=0$'') and $u_0=1$.
To test the robustness of this approach, we consider light curves with
$u_0=4,2,1,1/2,1/4,1/8$ and fit each of them to each of these two
forms, i.e., by adjusting $(t_0,t_\eff)$ to get the best fit.
Figure~\ref{fig:hm} shows that for $u_0=1/4$, the high-magnification
formula $A_1(Q)=Q^{-1/2}$ already works extremely well, while at
$u_0=1/8$, the input curve is almost indistinguishable from the model.
On the other hand, Figure~\ref{fig:lm} shows that the $u_0=2$ and
$u_0=4$ inputs are quite well fit by the $u_0=1$ (i.e., $A_2$) model.
Note that $u_0=4$ ($A_\max\sim 1.006$) is already at, or somewhat
beyond, the practical limits of microlensing detection.  The most
difficult case is $u_0=2$.  This differs from both model curves by a
noticeable amount, but these differences are still quite small.  Since
these are purely theoretical investigations, the conclusion that one
of $A_1$ or $A_2$ will fit the data acceptably regardless of the true
$u_0$ is independent of the magnitude of the event.

Next, we ask how $\Delta\chi^2$ is affected if the true value of $t_0$
differs from the model $t_0$ by $\epsilon t_\eff$.  Note that, by
construction, $\epsilon\leq\delta_{t_0}/2$.  We can quantify the
offset between the grid-based and true-model light curves by
$\delta\chi^2/\chi^2$, where $\delta\chi^2$ is the difference between
the best $\Delta\chi^2$ and the one that is obtained from a fit forced
to $u_0=0$ (or $u_0=1$), while $\chi^2\equiv\sum_i
[f_s(A_i-1)/\sigma_i]^2$.  Note that (because the algorithm compares
the observed lightcurve to its mean -- rather than a true baseline)
$\Delta\chi^2\sim\chi^2/3$.  Assuming uniform cadence and below-sky
errors, one finds that for high-magnification light curves
$A_1(Q)=Q^{-1/2}$,
\begin{equation}
{\delta\chi^2\over
\chi^2}\simeq 1 -
{\int_{-\infty}^\infty dx [(1 + (x-\epsilon/2)^2)(1 + (x+\epsilon/2)^2 )]^{-1/2}
\over\int_{-\infty}^\infty dx (1 + x^2)^{-1}}
\rightarrow {\epsilon^2\over 8}.
\label{eqn:dchi2}
\end{equation}
Since, as mentioned above, $\epsilon\leq\delta_{t_0}/2=1/6$,
this ratio is negligibly small.

The $\Delta\chi^2$ deficit due to discretely chosen $t_\eff$ leads to
a similar calculation with the result:  $\delta\chi^2/\chi^2=\epsilon^2/32$.

The calculations of this section show that our discretization choices
are, if anything, too conservative. That is, we could afford to use a
coarser grid.  We discuss in Section~\ref{sec:binary} why such
conservative choices are appropriate at this stage and why they may
possibly be relaxed in future years.

\section{Implementation for the KMTNet 2015 Data}
\label{sec:2015data}

In addition to introducing the algorithm itself, this paper has two
main goals.  The first, addressed in this section, is to identify the
strengths and weaknesses of the algorithm by applying it to a real
data set.  The second is to make available the results of the event
search to the general microlensing community, which we will address
further in Section \ref{sec:policy}.  Both of these goals require that
the algorithm be specifically applied to our data set through a series
of concrete choices.  One choice, for example, is whether to search
for events in the data from a single observatory or whether to conduct
the search by simultaneously fitting data from all three (or perhaps
just two) observatories.  Another choice is whether to initially vet
for variable stars and artifacts by some sort of algorithm or whether
to reject these totally in the operator stage ``by eye''.

The actual choices made for the 2015 data were strongly conditioned by
the operational environment, including both the nature of the 2015
data and the operational pressures affecting the reduction and
analysis of these data.  It is therefore necessary to summarize these
operational constraints to understand both the empirical evaluation of
the algorithm conducted below and the data products, as well as to
develop ideas on how to improve application of the algorithm in the
future.  The last point is discussed in Section~\ref{sec:improve}.

\subsection{Operational Constraints}
\label{sec:operations}

Much of the 2015 data is of very high scientific quality, and they have
already been the basis of many published and submitted scientific
papers \citep{ob151285,
ob151212,
isolated1,
ob150954,
ob150768,
ob150051,
galdist1,
%ob151459,and ill
ob151482}
However, 2015 was the commissioning year and as such it was heavily
impacted by engineering tests as well as physical problems,
refinements, and repairs.  For example, the electronics of the KMTC
camera were changed substantially on HJD$^\prime\equiv {\rm
  HJD}-2450000=7129$ and 7192, leading to discontinuities in the
photometric scale in some (but far from all) light curves at those
dates.  There were further experiments on the KMTC detector
electronics that sometimes resulted in quite noisy data.  For most of
2015, KMTS was impacted by condensation on one of the cooling lines,
which triggered a short in a wallboard (before the problem was
recognized) and so impacted photometry until it was replaced.  And
while KMTC and KMTS observations began very close to the beginning of
the season (HJD$^\prime\sim 7050$), KMTA began only on
HJD$^\prime=7180$.  There were many other issues as well, including
problems with the polar alignment and pointing model.
Problems of this order are completely normal for the commissioning of
relatively complex systems, but they also imply that algorithmic ideas
that are designed for basically homogeneous data must be significantly
adjusted if they are to function in this environment.

The first step was to write a pipeline to extract photometry from the
images.  This pipeline construction impacted the event-finder in
several ways, both direct and indirect.  This will be discussed in the
next several subsections.  However, from the present standpoint the
main impact was to delay the development and testing of the
event-finder.  It required about 8 months to develop and test the
pipeline and another 3 months, using all available computer resources,
to extract light curves just for KMTC.  Testing of the event-finder
could proceed in parallel with light curve extraction, but full
implementation could only be carried out after the light curves were
extracted.  Hence, event finding for 2015 did not begin in earnest
until the end of the 2016 season.

The combined effect of these considerations was to drive us toward
event-finding procedures that could be implemented quickly and would
be robust in the face of inhomogeneous data.  Work on more
sophisticated versions was deferred, although this was subsequently
initiated.  See Section~\ref{sec:improve} for a discussion of the
progress on a more advanced version.

\subsection{Catalog Construction and Pipeline}
\label{sec:pipeline}

The most straightforward way to extract light curves is to measure
photometry at the positions of known stars. This requires a catalog of
stars matched to the detector coordinate system.

The original approach that had been taken was to carry out
point-spread-function (PSF) photometry using the DoPhot package
\citep{dophot}, which extracts a catalog directly from the images
themselves. Then, those DoPhot light curves would be searched for
events, and finally difference image analysis (DIA, \citealt{alard98})
re-reductions using the \citet{albrow09} pySIS package would be used
for events of particular interest.  This plan was well-matched to the
available computer resources but suffered from what turned out to be a
fatal flaw: the DoPhot-based catalogs contained a factor $\sim 5$
times fewer entries than the number of potential source stars that
give rise to usable microlensing events.  If not corrected, this would
reduce the potential power of the experiment by roughly a factor
three\footnote{This factor is less than the naive ``five'' because,
for identical event geometries, brighter sources have smaller
magnitude errors, making it is easier to detect and measure subtle
anomalies.}.

A new pipeline was written based on publicly available DIA code from
\citet{wozniak2000}.  The central problem remained, however, of
constructing a deep input catalog.  While various packages were tried,
it proved impossible to even come close to matching the depth of the
existing OGLE-III \citep{oiiicat} catalog.  The reasons for this are
not completely clear.  No doubt, this is partly explained by the fact
that OGLE pixels are much smaller than KMTNet pixels
($0.26^{\prime\prime}$ versus $0.4^{\prime\prime}$), but it is
difficult to believe that this is the full explanation.  Probably the
biggest factor is just greater photometric experience.

The OGLE-III catalog provides stellar positions in RA and Dec, so it
was astrometrically matched to the template frame, i.e., the image
coordinates, by cross-matching with the DoPhot catalog. The pipeline
uses the OGLE-III catalog stars and photometric scale whenever
possible. However, the OGLE-III catalog does not cover the full extent
of the 2015 KMTNet fields. Some KMTNet fields cover fairly broad
continuous regions that are not covered by the OGLE-III survey.  In
addition, the OGLE-III catalog has non-trivial topology, due to gaps
of various sizes and shapes between OGLE-III fields.  Approximately
10.3\% of the area covered by the four prime KMTNet fields is not
covered by OGLE-III (see Section~\ref{sec:artifacts}).  In these
regions, the input catalog is derived from DoPhot templates and the
photometric scale is calibrated by cross-matching with the OGLE-III
catalog in nearby regions. In total, our OGLE-III + DoPhot catalog
consists of 71,555,640 stars.

\subsection{Automated Light Curve Search}
\label{sec:autosearch}

Once the light curves for all catalog stars are extracted, they must
be searched for microlensing events. Note that for 2015 data, only
KMTC light curves were used to search for events (as indeed these were
the only data reduced).  The basic algorithm for searching for the
best microlensing model was described in Section \ref{sec:algo}. Its
basic features were to perform linear fits to the data on a grid of
$(t_0, t_\eff, u_0)$ and to fit only those data from the interval
($t_0-5\,t_\eff<t<t_0+5\,t_\eff$). In practice, those linear fits were
performed in four phases. First, all the points were included in the
fits to evaluate the best-fit parameters, $(f_0,f_1)$.  Then, the
exact procedure was repeated to determine the naive $\chi^2$ of each
data point, using the $(f_0,f_1)$ from the previous step.  Then the
10\% worst-$\chi^2$ points were eliminated, and the data were refitted
to redetermine $(f_0,f_1)$. Finally, these $(f_0,f_1)$ were used to
determine the overall $\chi^2_{\rm \mu lens}$.

To determine the significance of the microlensing signal, the final
sample of points (i.e., with 10\% points removed) was then fitted to a
flat line, i.e., a constant flux level. Nominally, this would imply a
single parameter, i.e., the mean flux during the interval probed
($t_0-5\,t_\eff<t<t_0+5\,t_\eff$).  In fact, however, to allow for
breaks in the light curve due to firmware adjustments at KMTC, each of
the three intervals described in Section~\ref{sec:operations} was
permitted a separate mean flux level.  Hence, there could be one, two,
or three such parameters depending on how many of these three
intervals were included in ($t_0-5\,t_\eff<t<t_0+5\,t_\eff$).  This
procedure yields $\chi^2_{\rm flat}$.  Finally, the renormalized
$\Delta\chi^2$ was derived,
\begin{equation}
\Delta\chi^2 \equiv \biggl({\chi^2_{\rm \mu lens}\over\chi^2_{\rm flat}}
-1\biggr)N_{\rm data}
\end{equation}
where $N_{\rm data}$ is the number of data points remaining after
rejecting the worst 10\%.

\subsection{Automated Light Curve Grouping}
\label{sec:autogroup}

Several catalog stars can yield very similar light curves, i.e., with
similar $t_0$ and $t_\eff$.  If these are produced by stars in close
proximity, it may be assumed that there is only one true variable
(whether microlensing or otherwise), and the other variations are
simply echos of the real one.  For example, there can be two catalog
stars within $1^{\prime\prime}$. If one undergoes a microlensing event
then when DIA is applied to the other (with $1^{\prime\prime}$
displacement of its tapered aperture), it will also generate a
microlensing-like light curve.  In this case, we would like to examine
only the better curve, i.e., the one with higher $\Delta\chi^2$.

We find that in the case of bright variables (or, very occasionally,
bright microlensing events) these echo light curves can extend over
many tens of pixels.  We therefore develop a friends-of-friends
algorithm to apply to candidates found in the above search.  A
``friend'' is defined as a light curve from another star within 10
pixels ($4^{\prime\prime}$) and with $|t_{0,1} - t_{0,2}|<
0.4(t_{\eff,1} + t_{\eff,2})$.  Then friends of friends are grouped
and only the highest $\Delta\chi^2$ member of the group is shown to
the operator.  While most groups are small, it is not uncommon to have
groups of several dozens, or even more than 100.

\subsection{Manual Light Curve Review}
\label{sec:manureview}

\begin{figure}
\plotone{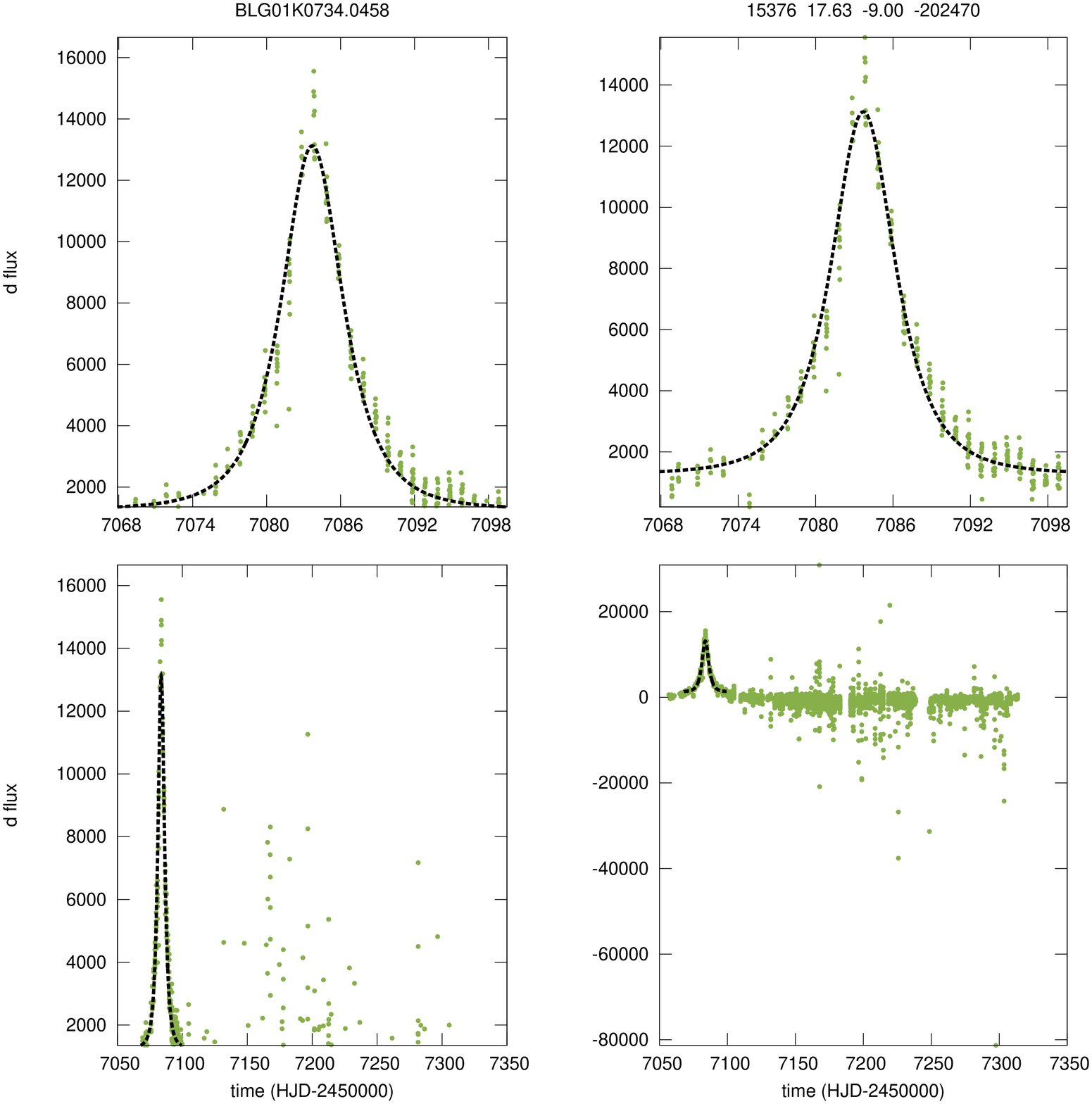}
\caption{Example of candidate light curve (ultimately judged to be
``clear microlensing'') as initially shown to the operator.  Top
panels show only the $10\,t_\eff\sim 30\,{\rm days}$ that were
fitted to the model (dashed curve), while bottom panels show full
season.  Left panels show only data in the range suggested by the
model curve (with small buffers at top and bottom) while right
panels show the full range of data.  Four numbers above the upper
right panel are $\Delta\chi^2$, $I_{\rm base}$, variability in
magnitudes, and variability in flux.  Since the last two numbers are
derived from the OGLE-III-catalog ``rms'' column, while this event
comes from an area not covered by OGLE-III, these entries are
negative, meaning: ``ignore''.
}
\label{fig:BLG01K0734.0458.jpg}
\end{figure}

\begin{figure}
\plotone{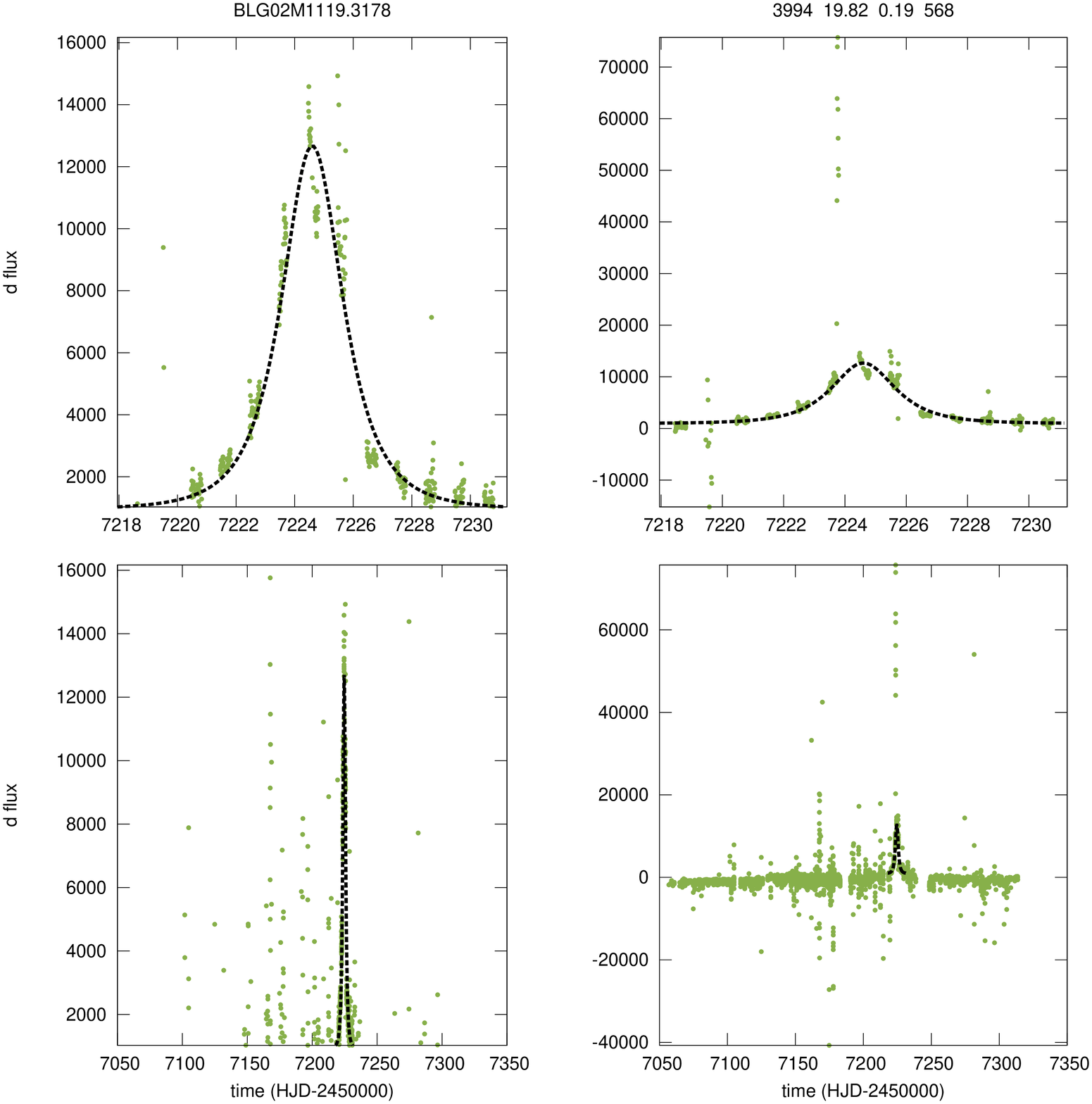}
\caption{Example of candidate light curve (ultimately judged to be
binary ``clear microlensing'') as initially shown to the operator.
Panel displays are the same as in Figure~\ref{fig:BLG01K0734.0458.jpg}.
In this case, however, the upper right panel is crucial to realizing
that the apparent ``noise'' near the top of the model in the upper-left
panel, is actually due to a strong binary caustic entrance.  Also note
that since the source star is in the OGLE-III catalog, the
OGLE-III ``rms'' (0.19 mag) is reported, and this is translated
into an estimate of 568 flux units in the figures.
}
\label{fig:BLG02M1119.3178.jpg}
\end{figure}

\begin{figure}
\plotone{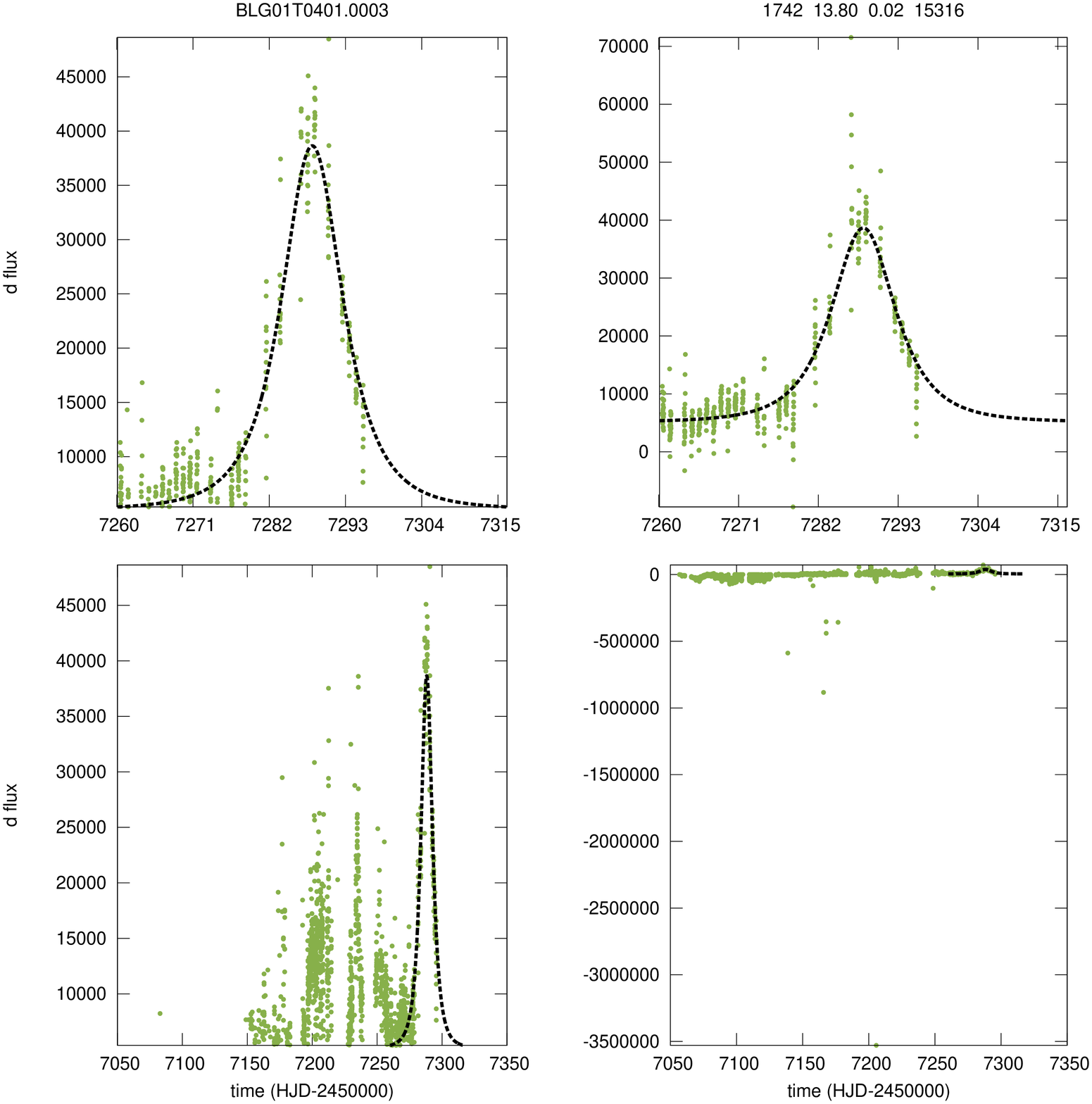}
\caption{Example of candidate light curve (ultimately judged to be
to be a ``variable'')  as initially shown to the operator.  The
upper left panel looks plausibly like microlensing.  However, it
is immediately obvious from the lower-left panel that there are
several other variations during the season of similar duration,
albeit of somewhat lower amplitude.  The assessment of ``variable''
is further confirmed by the OGLE-III based ``rms'' of 15316 flux
units.  This is only slightly smaller than the rms one would
measure from the ``microlensing event'' that is modeled in the
upper-left panel.
}
\label{fig:BLG01T0401.0003.jpg}
\end{figure}

\begin{figure}
\plotone{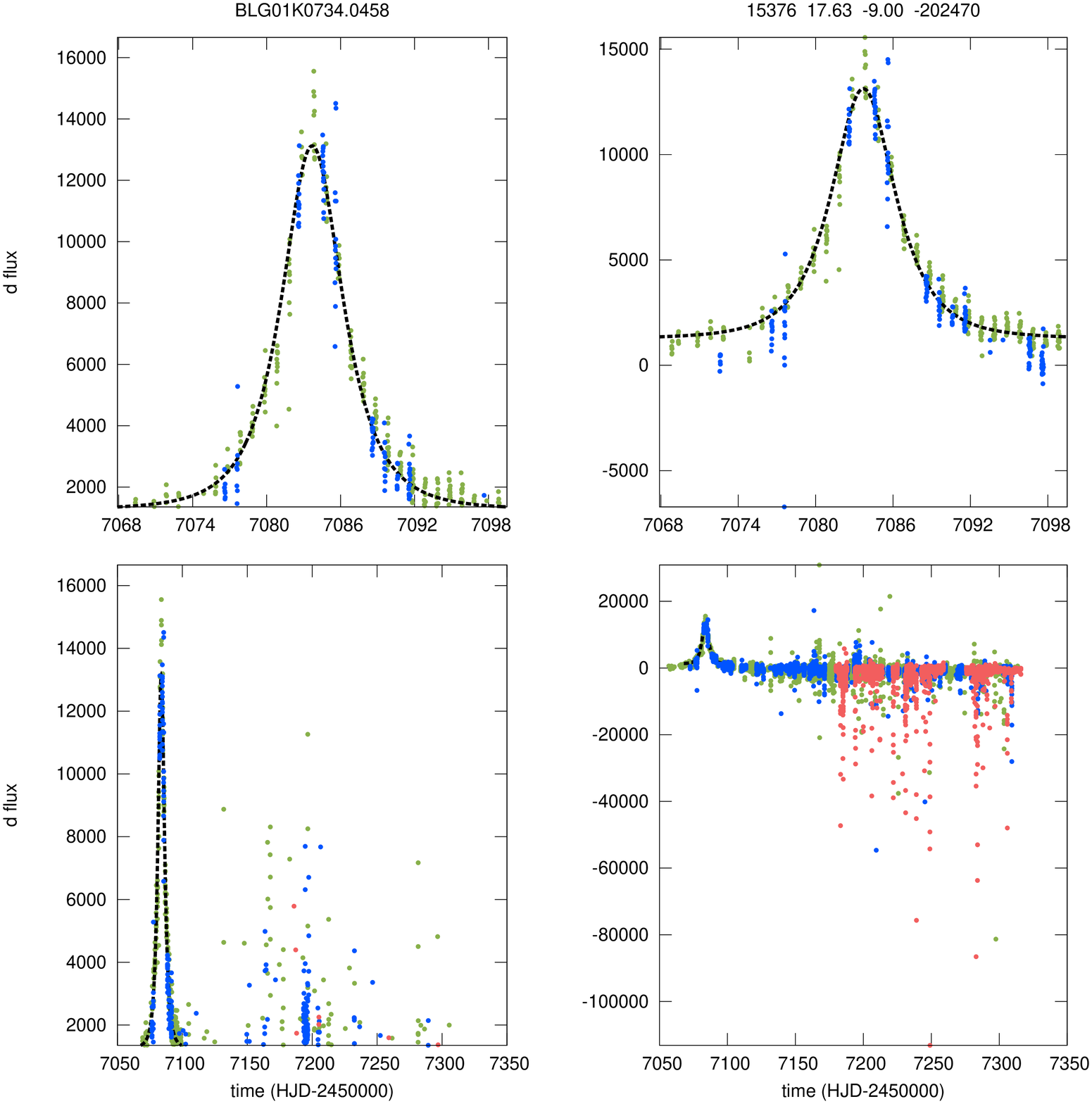}
\caption{Example of three-observatory light curve that the operator
reviewed after the same event (shown in Figure~\ref{fig:BLG01K0734.0458.jpg})
was judged to be either ``clear'' or ``possible'' microlensing.
The SAAO data (blue) confirm the microlensing character previously
indicated by the CTIO data (green).  The SSO data start too late in the
season to serve as a check in this particular case.  The operator would
also have been shown additional 4-panel displays for each observatory
separately at this stage.  Based on this inspection, the event was
judged to be ``clear microlensing''.
}
\label{BLG01K0734.0458_all.jpg}
\end{figure}

\begin{figure}
\plotone{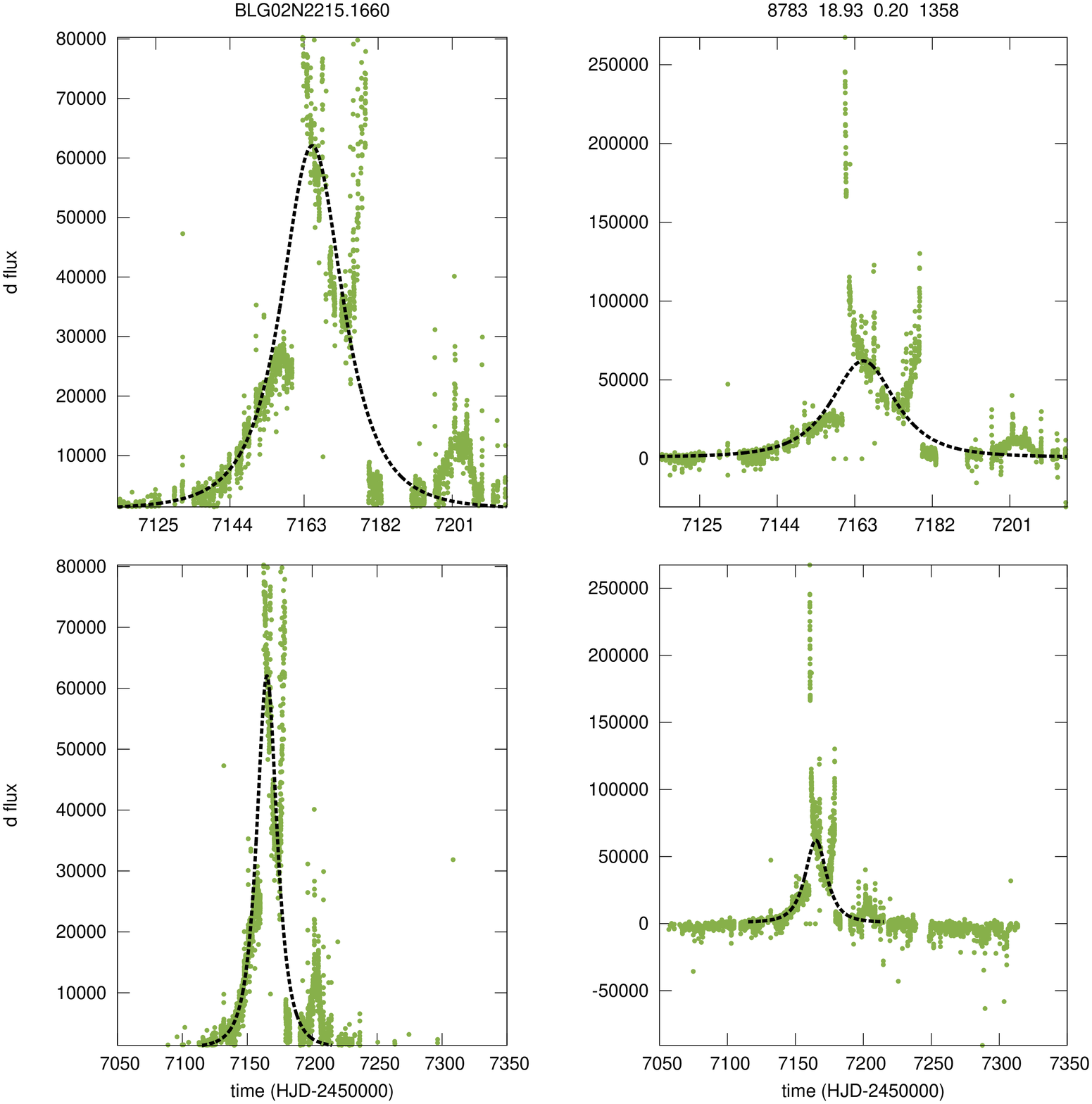}
\caption{Example of candidate light curve (ultimately judged to be
binary ``clear microlensing'').  Although the event bears little
resemblance to the form of point-lens microlensing (upon which
the event-finder algorithm is based), it is easily selected
by the algorithm $(\Delta\chi^2=8783)$ to be shown to the operator,
who in turn easily recognized it as binary microlensing.  In fact,
very few recognizable binaries are rejected by the algorithm.
See text.
}
\label{fig:BLG02N2215.1660.jpg}
\end{figure}

If $\Delta\chi^2>500$ then the candidate ``event'' is shown to the
operator in a four-panel display, together with some auxiliary
information
(Figures~\ref{fig:BLG01K0734.0458.jpg}--\ref{BLG01K0734.0458_all.jpg}
and Figure~\ref{fig:BLG02N2215.1660.jpg}).
Each panel contains all the data (not
just the data surviving the 10\% rejection). One reason to show all of
the data is that caustic crossings in binaries are likely to be
removed by the 10\% rejection cut as ``outliers'' from the point lens
fit.  The bottom two panels
show the whole season of data, while the top two show only the
$10\,t_\eff$ that were fitted to the model.  The left two panels are
restricted to the flux range predicted by the model (plus small border
regions at top and bottom), while the right two panels show the full
range of data values within the temporal range of the diagram.

These four displays allow the operator to rapidly determine whether
this is plausibly a microlensing event.  For example, the upper left
panel gives a very clear impression of whether the event is reasonably
well fit by point lens microlensing over the $10\,t_\eff$ range that
is being fit (see Figure~\ref{fig:BLG01K0734.0458.jpg}).  If, on the
other hand, the event has pronounced binary structure (e.g., a
pronounced caustic near the peak of the event), then this structure
will mostly not be visible on this plot, and the light curve may
appear to deviate from point-lens microlensing in an unfamiliar way.
However, this caustic structure will appear clearly on the upper right
panel, even if the main part of the microlensing event (easily visible
on the upper left) now is so small that it is barely visible (see
Figure~\ref{fig:BLG02M1119.3178.jpg}).

This approach also permits rapid rejection of variables.
The
algorithm itself quite easily fits wide classes of variables
to microlensing light curves because it ``censors'' data from
outside the $10\,t_\eff$ fitting interval. However, this is
readily apparent from the two full-season light curve panels
(see Figure~\ref{fig:BLG01T0401.0003.jpg}).

Above the upper-right panel are displayed four numbers
$(\Delta\chi^2, I_{\rm cat}, {\rm RMS(mag)},{\rm RMS(flux)})$.
Here $I_{\rm cat}$ is the magnitude of the catalog star,
RMS(mag) is the amplitude of scatter of this star derived
from the OGLE-III catalog, and RMS(flux)
$\equiv {\rm RMS(mag)}\times 10^{0.4(28-I_{\rm cat})}$.
The last number then gives the expected variability in flux units.
For OGLE-III stars, $I_{\rm cat}$ is derived from the
OGLE-III catalog.  For non-OGLE-III stars, it is derived by
calibrating the DoPhot photometry based on the overlap with OGLE-III
areas.  In this case, there is no information on
variability, so the final two numbers are set to arbitrary negative
numbers, meaning ``ignore''.

The operator then classifies the candidate into one of four
categories: ``clear microlensing'', ``possible microlensing'',
``variable'', and ``artifact''.

\subsection{Three-Observatory Confirmation}
\label{sec:confirm}

All ``clear'' or ``possible'' microlensing events are then further
evaluated using KMTS and KMTA data.  As mentioned above, for 2015,
only the KMTC data were reduced en masse. For candidate events, KMTS
and KMTA light curves were extracted from individual $(256\times 256)$
patches cut from each image taken by these two observatories, which
were then analyzed using DIA to produce light curves.  These light
curves are then sent through exactly the same ``event-finder'', except
with a threshold $\Delta\chi^2>0$ (instead of $\Delta\chi^2>500$) so
that some result will be reported regardless of the quality of the
fit.  The operator then views all three individual displays (each with
four panels) and a display with the combined data to make a final
decision, again choosing from the same four categories (see
Figure~\ref{BLG01K0734.0458_all.jpg}).

For genuine microlensing events, the most usual situation is that the
other two observatories (or, for events before the onset of KMTA, one)
have independently observed the same event and the event-finder has
``found'' this event and appropriately displayed it.  In a few cases,
the event-finder will find a ``better candidate'' in one of the data
sets that is actually an artifact. Still it is relatively easy to
check from the two full-season displays and the combined plot that the
real event is present in the data.  In these cases, the event can be
fully confirmed as ``clear'' even though its KMTC-only classification
is ``possible microlensing''.  There are some instances in which the
other two observatories lack data during the candidate event's time of
significant magnification, in which case the classification must be
based on KMTC data alone.

We note, however, that even when there are data from several sites,
these do not constitute absolutely fool-proof security against false
positives.  First, eruptions by cataclysmic variables (CVs) can mimic
microlensing events.  Having data from several sites can help guard
against these, particularly because CVs usually rise faster than
microlensing events (for the same fall time) and sometimes these rises
are not present in KMTC data but are present in KMTS or KMTA.
However, in other cases, it can remain difficult to distinguish,
either because there are few/no rising data or because the CV rises
unusually slowly.

One of the biggest problems in identifying microlensing events from a
single year of data is that there are no ``out of year'' baseline data
with which to reject variables.  For short events, this is not much of
a problem because there are ``in year'' baseline data.  However, there
is no unambiguous way to distinguish full-season microlensing events
from long-period variables.  In some cases, particularly for the
catalog stars derived from OGLE-III (the overwhelming majority), the
variables can be vetted using ``RMS(flux)''.  However, this is not
completely secure, since variables can remain dormant for several
years or vary more in a given year than in several previous years.
For the cases for which there is no obvious reason to reject the
microlensing explanation, these are called ``possible microlensing''.

\subsection{Vetting for Artifacts by Image Inspection}
\label{sec:artifacts}

A less obvious path to ``false positive confirmation'' comes from artifacts.
Based on the commissioning issues described in
Section~\ref{sec:operations}, one can well imagine that there are a
large number of artifacts in KMTC data that trigger the event finder.
In the overwhelming majority of cases, these are easily recognized as
such (or perhaps ``misclassified'' as variables) and so do not make it
to the stage of vetting by KMTS and KMTA comparisons.  In many of the
remaining cases, these artifacts are easily removed because they are
not duplicated in data from one or both of the other observatories.
However, a large fraction of candidates initially classified as
``possible microlensing'' and a handful of those initially classified
as ``clear microlensing'', turn out to be examples of one of two types
artifacts: ``bleeding columns'' and ``displaced variables''.  We
developed simple procedures for identifying these.

For each of the events that was classified as either ``clear'' or
``possible'' microlensing, we display side-by-side the finding chart
and a difference image from near the peak of the ``event''.  Below
these are the set of light-curve displays described in
Section~\ref{sec:confirm}, and below these panels are a set of four
difference images: the two nearest the peak, and two other good seeing
images relatively near peak.

We place cross-hairs at the catalog position of the source in all
images.  From the top-level difference image, one can easily see
whether or not the cross hairs are at the position of the variation.
If variation is not at the catalog (cross hair) position, this is not
in itself a problem: many microlensing events occur on stars that are
too faint to be cataloged, and are recognized by the flux variations
that they induce at the position of a neighboring catalog star due to
the finite width of the PSF.  However, if the finding chart contains a
bright star at the position corresponding to the flux variation on the
difference image, then the ``event'' is almost certainly an artifact,
i.e., an echo of variations in this bright star.  Of course, a given
bright star might actually undergo a microlensing event, and this
event would likewise be ``echoed'' at the positions of neighboring
cataloged stars.  But in this case we would expect that the bright
star would itself be cataloged and would yield a microlensing light
curve with higher $\Delta\chi^2$.  For variable sources, however, the
``cleaner'' light curve from the variable is much more easily excluded
as a ``variable'' than is the ``echo'', so only the echo winds up in
the list of candidates.

Moreover, it is almost always the case that these echos last for most
or all of the season, simply because repeating variations would have
been recognized as variables, even in their echos.  In the course of
this vetting, we noted a handful (2--3) cases for which the source
position was offset and coincident with a bright star, but the event
was nevertheless short and well contained within the season.  We
accepted these as microlensing events and assumed that either the
catalog or the event finder had failed to function properly in these
cases.  This means that there were probably also a few long
microlensing events with similar characteristics that we misclassified
as variables, but this problem cannot be addressed with only a single
year of data.

Another quite common artifact that easily shows up in the
difference images are fake events due to bleeding columns.
In fact, the level of bleeding is far too low to be noticed
directly in the images, and can only be perceived in difference
images, which permit a much stronger ``stretch''.  Time variable
bleeds can be generated by variable stars, in which case they are
typically either long-timescale or repeating.  However, they
can also be generated by shorter-period irregular variables
that only rise above the bleed threshold for a relatively short
period, once in a season.  In addition, they can be caused by
changes in seeing and background.  The former is not likely
to be correlated among observatories, but the latter is, i.e.,
if it is due to lunar phase.  In any case, as mentioned above,
such bleeding columns are easily recognized by examining the
four images that we routinely display.

\begin{figure}
\plotone{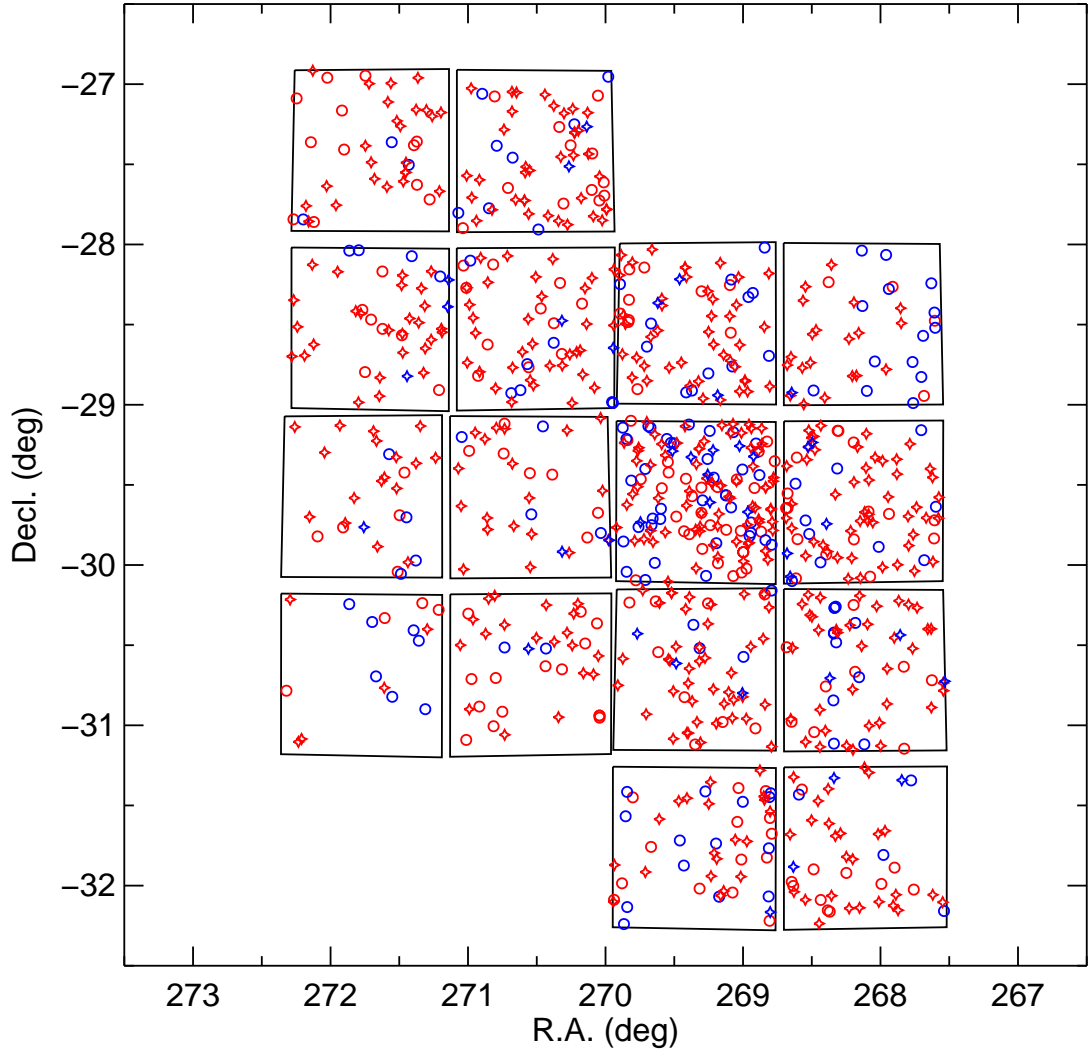}
\caption{Four fields observed by KMTNet in 2015, BLG01, BLG02, BLG03, BLG04
each with four $1\,{\rm deg}^2$ chips, T, K, M, N
(both counterclockwise from lower right).  Red and blue indicate
``clear'' and ``possible'' microlensing.  Star symbols are events
found previously by OGLE and/or MOA, while circles were not
previously alerted.
}
\label{fig:geom}
\end{figure}

The event finder yielded 955,659 candidates, which were automatically
grouped into 385,565 groups to be shown to the operator.  Of these,
148,010 were classified as ``variables'', and 236,698 as ``artifacts''.
Our final catalog of events consists of 673 ``clear microlensing'' 
and 184 ``possible microlensing'' events (with some duplicates).  
See Figure~\ref{fig:geom}. 
Section \ref{sec:esum} summarizes the properties of our final sample. 
 
\begin{figure}
\plotone{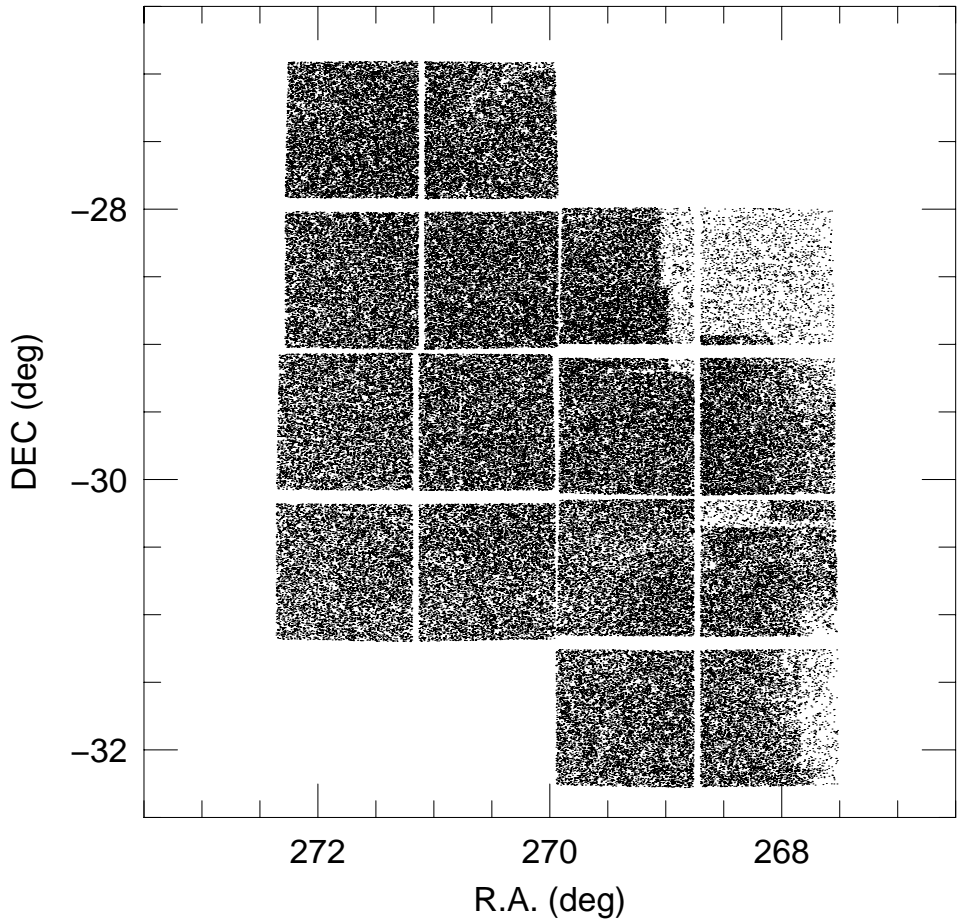}
\caption{Star catalog density of
four fields observed by KMTNet in 2015, BLG01, BLG02, BLG03, BLG04
each with four $1\,{\rm deg}^2$ chips, T, K, M, N
(both counterclockwise from lower right).  One per 500 catalog stars
is plotted.  Sharp rectangular boundaries in density are due to
regions not covered by the OGLE-III catalog, where DoPhot finds only
about 1/5 as many stars.  Note that the catalog density is only about
1.35 times higher in BLG02N compared to BLG04N, but that there are
an order of magnitude more ``clear microlensing'' events.  
See Figure~\ref{fig:geom}.
}
\label{fig:fldmap}
\end{figure}

It is interesting to compare this map of detections with a map of the
underlying catalog (Figure~\ref{fig:fldmap}).  Note in particular that
BLG02N has a density only about 1.35 times higher than BLG04N but has
an order of magnitude more ``clear microlensing'' events.  This is
broadly consistent with our expectations based on \citet{poleski16},
which shows that the microlensing event rate should be highest near
the center of BLG02N.

\subsection{Comparison to OGLE-IV EWS Detections}
\label{sec:comp}

The best external check on the event-detection procedures described
above is to compare with OGLE-IV. Since OGLE operates from a
very nearby site in Chile, it has very similar field visibility
and weather to KMTC, i.e., the KMTNet site used for the initial selections.
In fact, the visibility is not identical because OGLE observes
to greater hour angle than KMTC and also longer into austral spring.
On the other hand, KMTNet observes during all lunar phases, while
OGLE typically halts observations when the Moon is in the Galactic Bulge.
Similarly, KMTNet observes in all atmospheric conditions, provided
that these do not endanger the telescope and provided that the sky
is not opaque.  Further, for about a quarter of the KMTNet prime area,
OGLE-IV observes with cadence $\Gamma=3\,{\rm hr}^{-1}$, which
is not qualitatively different from the KMTNet cadence.  And, for most of
the remaining area, OGLE-IV observes with cadence $\Gamma=1\,{\rm hr}^{-1}$,
which still should be sufficient to detect a substantial majority
of microlensing events that are accessible to KMTNet. 

For this comparison, we use the set of OGLE-IV events announced by
their Early Warning System \citep[EWS][]{ews2}. These events are
identified in real-time based on only a partial light curve rather
than a full microlensing fit to a completed event. As such, this
comparison sample may not be complete and it may also contain some
false positives. However, it represents the most complete, publicly
available, independent list of microlensing discoveries for the 2015
season.

\subsubsection{{OGLE-IV Events Not found by KMTNet
}}

428 OGLE-IV events (after eliminating duplicates) in the KMTNet field
were not recovered by our event finding procedure, neither as
``clear'' nor ``possible'' microlensing events.  These represent
potential false negatives and may indicate problems with our
procedure. We first assess the fraction of the OGLE-IV alerts
discovered as a function of brightness. We define $I_{\rm peak} = 18 -
2.5*\log [f_{\rm S}(A_{\rm max} - 1) ]$, where $f_{\rm S}$ and $A_{\rm
  max}$ are determined from the OGLE EWS table. For the following
ranges of $I_{\rm peak}$, we find the recovery percentages are
$I_{\rm peak} (< 14\, $mag$\, = 63\%$, %0.6341463414634146 26 41
$[14$--$15] = 91\%$, %0.9090909090909091 10 11
$[15$--$16] = 76\%$, %0.7647058823529411 26 34
$[16$--$17] = 72\%$, %0.7166666666666667 43 60
$[17$--$18] = 72\%$, %0.722972972972973 107 148
$[18$--$19] = 61\%$, %0.6052631578947368 138 228
$[19$--$20] = 43\%$, %0.4291044776119403 115 268
and $>20 =  16\%$), %0.16260162601626016 20 123
i.e., the events missed by our procedure tend to be biased toward the faint end.

To explore this in more detail,
we examined one out of each 20 OGLE-IV events in the
KMTNet fields that were not found by the procedures outlined above.  This test
was conducted after selection based on KMTC data,
and before vetting based on KMTS and KMTA data, in order to provide
the closest basis of comparison.  One result of this test was that
we found that three of the 22 tested events lay in image patches
that had not been processed by the
event-finder due to a failure in the computer architecture (and
so having nothing to do with the event finder itself).  These computer
problems were fixed and these patches were re-run.  We report
only on the remaining 19 tested events.

Of these 19 ``missing events'', three were clear failures, three were
judged to be ``marginally acceptable'' false negatives, and remaining
13 were ``acceptable'' false negatives.

Two of the clear failures (OGLE-2015-BLG-0693 and OGLE-2015-BLG-1589)
were due to operator error.  The first was mislabeled as ``CV'', and
the second should have been called ``possible microlensing'' (and then
further investigated using KMTS and KMTA data).  The third failure
(OGLE-2015-BLG-1015) lay in a thin band between two patches.  
Each $1\,{\rm deg}^2$ chip of the 2015
data was analyzed in a ($40\times 40$) grid of ``overlapping''
($256\times 256$) patches, but in a few cases (given the ``dead zone''
boundaries imposed by the DIA software) the patches did not actually
overlap.  This issue is resolved for 2016+ data by going to much
larger patches with much larger boundaries.

Two of the three ``marginally acceptable'' cases (OGLE-2015-BLG-0870
and OGLE-2015-BLG-1138) were missed because the operator judged the
data to be too noisy for reliable microlensing detection.  In
hindsight (and knowing that there is a real event there),
these might have been classified as ``possible'' and further
investigated using KMTS and KMTA data.  However, these events were
quite near the threshold at which ``letting in'' additional events
would have led to vast increase in ``investigations'' of pure noise
and variables.

The third marginal case (OGLE-2015-BLG-1589) was not shown to the
operator because $\Delta\chi^2=436$.  If it had been shown, it
probably would have been classified as ``possible microlensing''.
Rectification of this ``problem'' would require setting the threshold
lower and so greatly increasing the already large number of events to
be vetted.  

\begin{figure}
\plotone{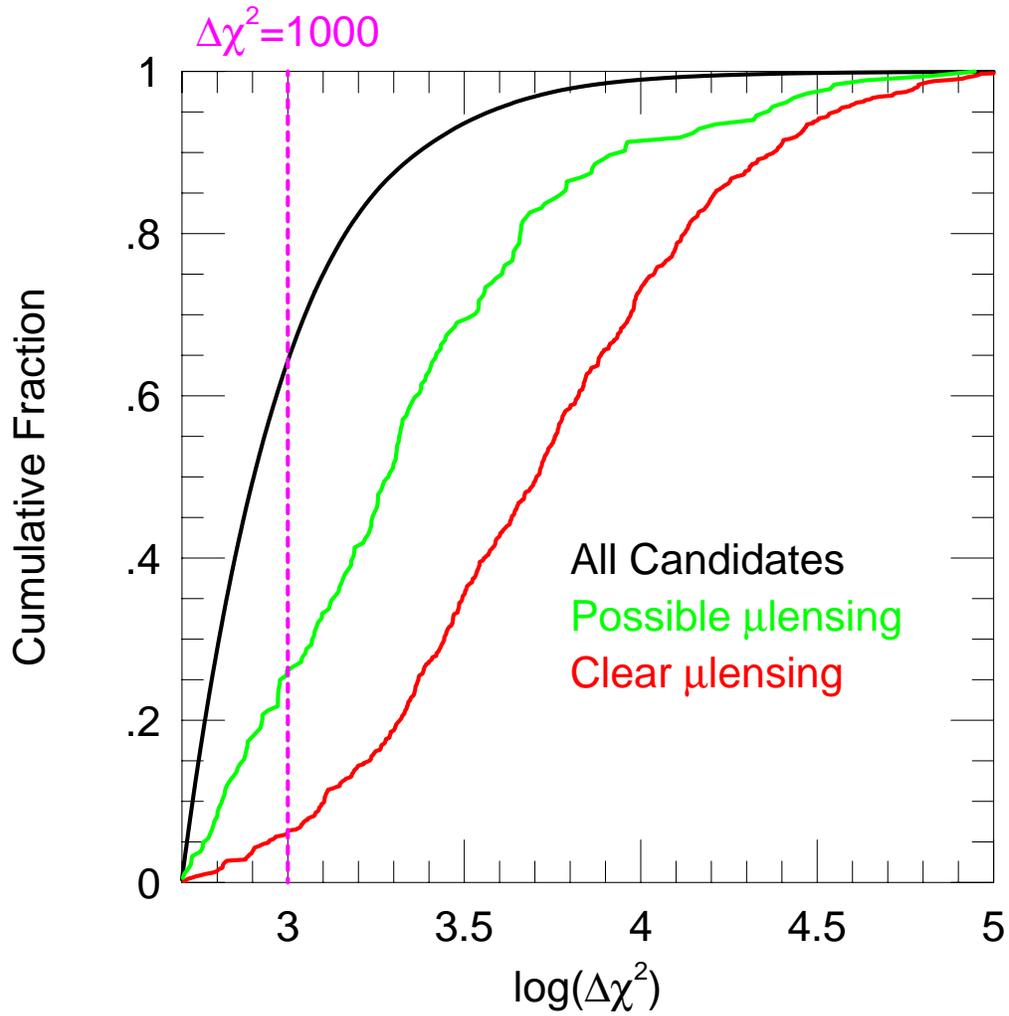}
\caption{Cumulative distribution functions in $\log(\Delta\chi^2)$ for
all 385,565 event groups (black), the 660 ``clear microlensing'' events (red)
and the 182 ``possible microlensing'' events (green).  Almost 2/3 of
the event groups have $500<\Delta\chi^2<1000$ (left of dashed line)
but these yield only
6.2\% of the ``clear microlensing'' events.
}
\label{fig:cum}
\end{figure}

To better understand the issues involved in possibly setting a lower
threshold, we show in Figure~\ref{fig:cum} cumulative distributions by
$\Delta\chi^2$ of i) all event groups, ii) ``clear microlensing''
events, and iii) ``possible microlensing'' events. Of particular note:
while 64\% of all (385,565) event groups had $500<\Delta\chi^2<1000$,
only 6.2\% of ``clear microlensing'' events lay in this range.

The remaining 13 event non-detections were all judged to be
``acceptable'' false negatives.  For two of these, (OGLE-2015-BLG-0059
and OGLE-2015-BLG-0960) the nearest stars in the catalog were
$1.2^{\prime\prime}$ and $2.4^{\prime\prime}$, which led to very weak
and no signal respectively.  Neither passed the $\Delta\chi^2=500$
threshold, but if they had, neither would have been chosen by the
operator.  Both events lie in regions not covered by the OGLE-III
catalog, and for which we are dependent on the much shallower DoPhot
catalogs.

One event peaked midway between 2014 and 2015 seasons, and so was not
recognizable based on 2015 data.  Two were long events that were
(correctly) judged by the operator to be consistent with being
variables based on 2015 data.  Two events were affected by artifacts
in the 2015 data.  The remaining eight had low S/N and so were not
shown to the operator, but if they had been they would have been
judged as too noisy for reasonable identification as microlensing
events.  (One of these, OGLE-2015-BLG-1835, may be a CV.)

In brief, this test shows that of order $3\times 20=60$ genuine
microlensing events that could have been detected in this region were
not.  These failures are mostly due to ``operator fatigue'' from
reviewing hundreds of thousands of light curves. In future years, this
may be ameliorated by eliminating most variables in advance and by
reduction in the number of artifacts now that commissioning is
complete.

\subsubsection{Events in KMTNet Not Found by OGLE-IV}
\label{sec:notOGLE}

For ``clear microlensing" events found by KMTNet but not found by the
OGLE-IV EWS, we would like to try to understand why they were not
detected by OGLE-IV.  For these events there are three possibilities:
OGLE could not detect (or would have extreme difficulty detecting) the
event due to insufficient data, OGLE excluded the event as a false
positive, or the OGLE EWS missed the event. Since OGLE has a different
camera, a different site, and different observing protocols, the data
on a particular event might be insufficient due to the location of
chip gaps or observability gaps. The OGLE survey has also been in
operation for many years and so can identify (and exclude) false
positives due to long-timescale, sporadic variables. We examine a
representative sample of the KMTNet events not found by the OGLE EWS
in greater detail to assess whether or not we expect OGLE to have
detected them. If it appears that OGLE ought to have detected an
event, then that event is a candidate for being a false positive,
i.e. it is possible OGLE did not alert the event because they had
reason to believe it was a variable star rather than microlensing. At
the same time, the absence of an alert is not conclusive evidence of a
false positive; the event could simply have been missed. On the other
hand, if the synoptic observing information suggests that OGLE would
have difficulty detecting an event, than the absence of an alert does
not provide any information about whether or not an event is a false
positive or real microlensing.

An unpublished study by Gould \& Udalski has already determined that
events detected by MOA but not by OGLE are overwhelmingly undetectable
by OGLE because they are in chip gaps, outside the OGLE fields, during
gaps in the data, etc. A few are variable stars that have been masked
from the OGLE catalog (in this case, two of 40, P. Mr\'oz, 2017,
private communication). Thus, if an event is detected by MOA but not
by OGLE, we do not investigate it further because there is likely a
plausible explanation as to why it was not alerted. We then reviewed a
random 10\% (i.e., 18) of the 177 remaining events to assess whether
or not there was sufficient data for OGLE to alert them.

We find that four of these randomly chosen 18 could not plausibly have been
detected by OGLE since they were either in chip gaps or
would have had very few magnified points
(BLG01N.3928.2251,
BLG02M.0326.3862,
BLG02T.0123.2655,
BLG03M.0939.0577).
Four additional events were very likely missed either because they
lay very near a chip edge (though formally within the chip) or
because they would have had only 3 or 4 significantly magnified points.
(BLG01T.2412.0089,
BLG02N.2218.3192,
BLG03K.3511.3254,
BLG04N.3638 2631). Thus, it is not possible to draw conclusions 
based the absence of an OGLE alert for these $4+4=8$ events.

There were four other events that appeared to us to be probably
detectable, but because they contained only 6--8 magnified points,
could have plausibly been missed, perhaps due to mildly adverse
conditions
(BLG02T.2218.3029,
BLG03N.2123.2010,
BLG04K.2626.3153,
BLG04T.2822.2817).  Here we should keep in mind that by selecting
non-OGLE events, we are biased toward such adverse circumstances.

Finally, there were six events that we thought should clearly
have been detected by OGLE
(BLG01K.0126.0793, % 9.99, offset src/nt-on-finder, flatbase, almost cmplt cov
BLG01M.1125.1711, % 9.99, offset src/pretty faint, flatbase
BLG01N.0633.3015, % 7.49, offset src/nt-on-finder, flatbase
BLG02N.0726.2197, % 1.33,I=19.4, flatbase
BLG02N.3209.0444, % 9.99 I=17.7, flatbase
BLG03T.1911.1388). % 1.33, offset src/nt-on-finder, flatbase
Five of these six lie in OGLE fields BLG505, BLG506, or BLG512, 
which had OGLE cadences of $\Gamma=3\,{\rm hr}^{-1}$ in 2015.
If there are false positives in our sample, these should be good
candidates.  In reviewing these events, we find that all look
plausibly like microlensing.  In all cases the $t_\eff<10\,$days and
the remainder of the light curve is consistent with being flat.  In
all cases, all three data sets basically agree on the form of the
light curve (except for the first, for which there are no KMTA data).
In one case (BLG01M.1125.1711) the variation is offset from the
catalog star and at the position of another resolved star, which we
would generally regard as a warning sign that the resolved star is a
variable and the ``event'' is an echo.  However, as discussed in
Section~\ref{sec:artifacts}, we override this concern when the event
is of short duration and the remainder of the season is flat.  This is
particularly true in the present case, for which the star that is
varying is relatively faint.  Hence, while any of these candidates
could in principle be false positives, there is no evidence that this
is the case. 

Subsequent to the posting of this paper on arXiv,
P. Mr\'oz (2017, private communication) kindly provided a list
of all events in OGLE high cadence fields that were recognized
as variables by OGLE.  This list contained one of the four we 
thought ``likely missed" by OGLE, two of the four events that 
we considered ``probably" should have been detected, but none 
of the six events that we considered should ``definitely" have been detected.

\subsection{Binary Detection}
\label{sec:binary}

We expect that our approach will be efficient at detecting planetary
events simply because these typically look similar to point-lens
events but with relatively brief anomalies.  One might be concerned
that these anomalies could reduce $\Delta\chi^2$ and so prevent their being
shown to the operator.  However, recall that the 10\% ``worst outliers''
to the point-lens-like fit are eliminated from the comparison with
a flat light curve.  Hence, we actually do not expect this to be an
issue.

The situation is, however, quite different for binary events,
many of which look nothing like point-lens events.  When we devised
the event-finder, we had no definite expectation of how well it
would perform on binaries.

We conducted a test to evaluate this performance after the fact.  We
manually examined all OGLE events lying in the KMTNet fields to find
those that either were, or plausibly could be binaries.  We found 57
binary events announced by the OGLE EWS that lie in the KMTNet
fields. Thirty-four are included in our list of microlensing events
(all of them are classified as ``clear". Of the 23 that were missed,
only one of these both plausibly looks like microlensing in KMTC data
and failed to be shown to the operator (i.e., had $\Delta\chi^2<500$).
This was OGLE-2015-BLG-0390, which shows a sharply rising caustic exit
in KMTC data but has $\Delta\chi^2=265$.  The reason for this low
$\Delta\chi^2$ is that the ``effective time scale'' of this rise is
much shorter than the 0.56 day minimum of the current search.  This
limit was in turn set by the large number of artifact-driven
``events'' that would be found in the 2015 data, which would have
greatly multiplied the work of the operator.

Of the other 22 that were missed, three were due to operator error:
one (OGLE-2015-BLG-0095) should have been classified as ``clear''
microlensing and two (OGLE-2015-BLG-1346 and OGLE-2015-BLG-2017) as
``possible microlensing''.  For the remaining 19, the signal in the
data (whether shown to the operator or not) was not sufficient to
count them as plausible microlensing events.  From the perspective of
understanding the algorithm, the key point is that only one known
binary was missed because the algorithm itself failed to detect it.

We consider it somewhat surprising that the algorithm is working
as well as it is on binaries, given that it is specifically
tailored to find point-lens events. One reason for this is that
the algorithm will detect pretty much any sort of bump provided
that it is sufficiently pronounced and not too sharp, as illustrated by
Figure~\ref{fig:BLG02N2215.1660.jpg}.
We suspect that the good binary
sensitivity then derives from the sampling density
of the 3-parameter point-lens grid described in Section~\ref{sec:robust}. While this sampling density is overly conservative for detecting point lens events, it likely permits matching with sharp caustic features, allowing the detection of microlensing binaries.  This conjecture
can be tested in the context of the improved event finder
(Section~\ref{sec:improve}) by repeating the fitting procedure on
known binaries with a progressively less dense grid. However,
because the event finding procedure is significantly less
computationally intensive than the photometry pipeline, there is no
strong driver to scale back the sampling density. If the computational situation changes, this test should be performed
before scaling back the sampling density.

\subsection{Summary of Detected Events}
\label{sec:esum}

In total, this procedure for identifying microlensing events from the
2015 KMTNet commissioning data resulted in 857 microlensing event
candidates; 15 of which are duplicates leaving 842 event
candidates. Of these, we classify 660 as ``clear" microlensing events
and 182 as ``possible" microlensing. Based on \citet{mroz15} and
P. Mr\'oz (2017, private communication), we find that 40 of the
``clear" microlensing events and 54 of the ``possible" microlensing
events were classified as variables by the OGLE collaboration. Of the
remaining events, 483 ``clear" events and 38 ``possible" events were
detected by either the OGLE EWS or the MOA alert system (or
both). Thus, KMTNet has detected 137 ``clear" microlensing events and
90 ``possible" events not found by any other survey. Based on our
investigation described in Section \ref{sec:notOGLE} and P. Mr\'oz
(2017, private communication), there is no evidence indicating these
are false positives.

\begin{figure}
\plottwo{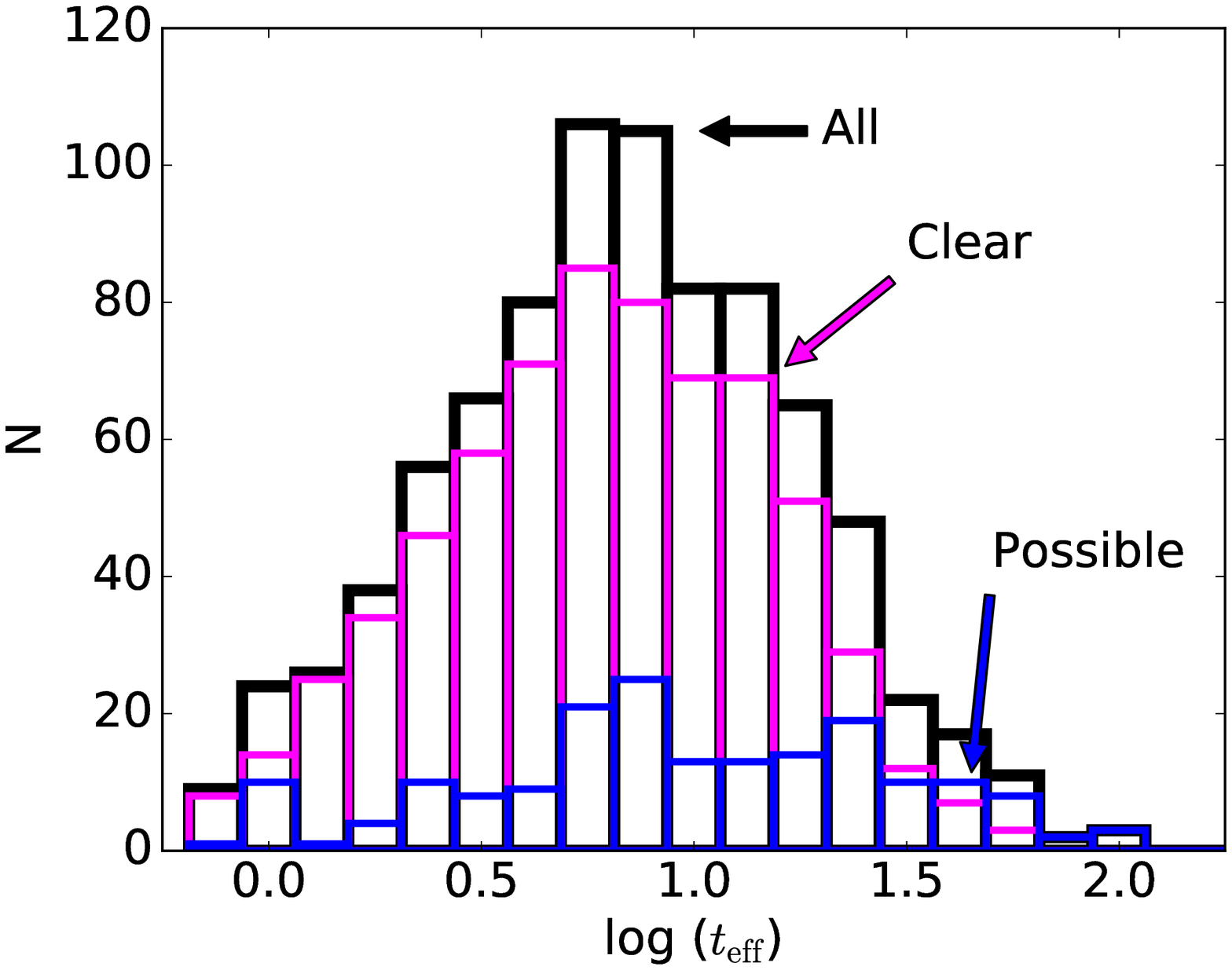}{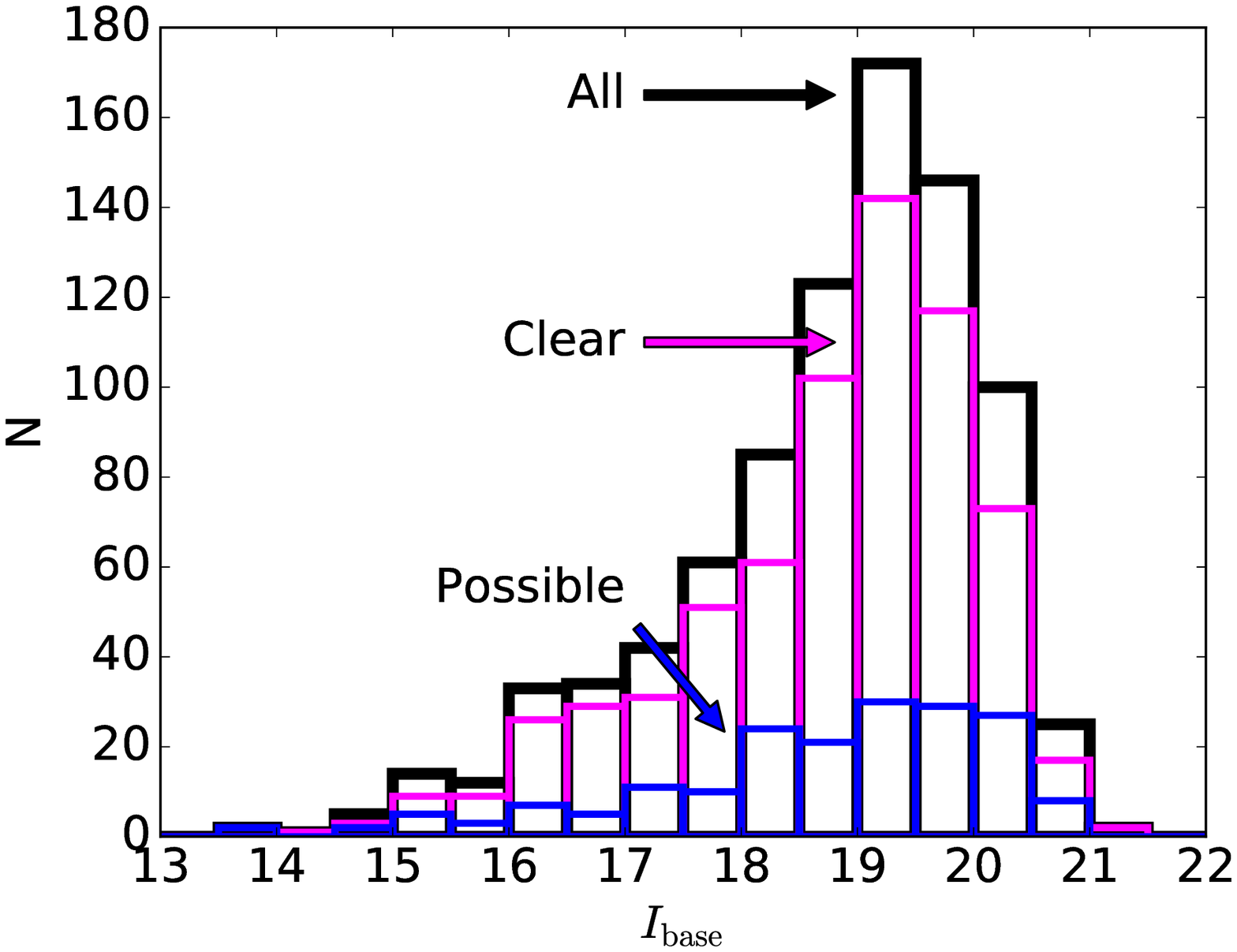}
\caption{The distribution of event detections as a function of $\log
  t_{\rm eff}$ (left) and $I_{\rm base}$ (right). The black solid
  line shows the histogram for all events; ``clear microlensing''
  events are shown as the magenta line, and ``possilbe
  microlensing'' events are shown as the blule line.}
\label{fig:hist}
\end{figure}

The distributions of $t_{\rm eff}$ and $I_{\rm base}$ for these events
are shown in Figure \ref{fig:hist}. The reader should bear in mind
that $t_{\rm eff}$ is discretely defined as described in Section
\ref{sec:algo} and therefore is not a perfect representation of the
event timescale.

\section{Data Policy and Data Releases} 
\label{sec:policy}

KMTNet data policy is framed by several goals, which are basically
compatible but are subject to some mutual tension.  First, we seek to give
adequate time for the KMTNet team to publish results based on the
huge amount of work required to obtain and process these data.
Second, we want the KMTNet data to be exploited to the maximum
extent possible, whether by KMTNet team members or others.  Third,
we want to promote the microlensing field by making data available
to as many workers as possible.  Fourth, we want to avoid overworking
our team on tasks that are auxiliary to the team's own effort
to publish our results.  Finally, we want to avoiding
infringing on the data rights of others, whether explicitly or
implicitly.

\subsection{KMTNet Website}
\label{sec:website}

KMTNet data are available at http://kmtnet.kasi.re.kr/$\sim$ulens/event/2015/ .
The main page is ordered by KMT event number in the form
KMT-2015-BLG-[NNNN].  These numbers are in turn sorted by field BLG[NN]
(01, 02, 03, 04),
chip (K,M,N,T), patch, and star number.  The ``clear microlensing'' events are
listed first, so that all entries $\leq 0673$ are ``clear'' and all
events $\geq 0674$ are ``possible''.  Each line contains the best
fit event-finder parameters $(t_0,t_\eff,u_0)$ (Equations~(\ref{eqn:general})
and (\ref{eqn:ajoftau})), the baseline flux, and the RA and Dec.
There are also cross references to OGLE and MOA discoveries.  By clicking
on the event name, one can see the finding chart as well as the same
16 panels of light curve displays shown to the operator.  These
pages also contain links to the light curve data.

\subsection{2015 Policy and Releases}
\label{sec:policy2015}

Guided by the above principles, we have formulated the following
policies for KMTNet data for 2015.  We then outline issues
that are under consideration for future years in the next section.

1) All 2015 KMTNet data remain proprietary until the acceptance
for publication of this paper and all papers based on 2015 KMTNet
data that are explicitly mentioned in Section~\ref{sec:operations}.

2) Beginning immediately (i.e., before the end of the proprietary
period), all light curves that are posted together with this paper
(see below) can be used by anyone to prepare future papers for
publication.  However, during the proprietary period (see (1)),
they cannot be submitted for publication, nor posted on arXiv.

3) Once the priority period ends (see (1)), papers can be submitted based
on the released data.  We welcome collaboration with the KMTNet
team, but do not require it as a condition for publication.
For the cases that we collaborate, the KMTNet team does not demand
any ``editorial control'', but rather simply reserves the right
to withdraw from papers with which we disagree strongly enough to
warrant such withdrawal.

4) If additional data processing is needed, we will either provide
re-reduced light curves in exchange for co-authorship or we will
provide flat-fielded ``stamps'' surrounding the event, provided that
sufficient evidence is given to us that there is substantive progress
toward publication of a paper.

5) No data will be given out for events discovered by other teams
but not independently discovered by KMTNet using the algorithm
presented here, unless there is explicit agreement with those teams.

The data that will be made available are

1) All the data used by the event finder for all events that were
determined to be either ``clear microlensing'' or ``possible microlensing''.
Recall that these data are based on \citet{wozniak2000} DIA using
input positions derived from a pre-constructed catalog.  Hence, they
are not usually of the highest quality, although sometimes they
are in fact close to optimal.

2) Automated \citet{albrow09} pySIS reductions of all light 
curves\footnote{Note that as of the time of submission,
these reductions are not yet complete, but will be uploaded
to the website as they are completed.}.
In roughly half of all cases, these reductions are better or substantially
better than the reductions from (1).  Unfortunately, in the other half
of cases, these automated reductions either basically fail or totally fail.
This means that, for a substantial fraction of events, obtaining very good
or excellent light curves requires additional, by hand, reductions.

No effort will be made to rectify these or any other residual problems for
the 2015 data release as a whole.  As discussed in Section~\ref{sec:operations},
the 2015 data have intrinsic problems that are not likely to propagate
into future years, and our main data reduction efforts will be applied
to those future-year data.

\subsection{2016+ Policy and Releases}
\label{sec:policy2016}

Future data policy and releases will be governed by the same principles,
but will be modified based both on the experience of the 2015 release
and the improved quality of 2016+ data.  For 2015, we are posting
light curves about 18 months after the close of the 2015 season.
We hope that this time lag will be substantially improved for 2016 and further
improved for 2017.  Since we do not know exactly what problems
we will encounter, we cannot guarantee it.  However, at this point
we believe that a 6-month delay is a plausible target for 2017+ data
because if a paper cannot
be drafted within 6 months of time the data are taken, it is likely
to fall down the list of priorities as the next season of data start
to become available.

For 2016, we hope to have an earlier release of the events in the
{\it Kepler} K2 C9 field.  (See \citealt{henderson16} for a description
of the K2 C9 project.)\ \
For KMTNet-discovered events, these data will have no
proprietary period.  We will also have a public release of KMTNet
data on events discovered by other teams but not discovered by KMTNet.
However, for
these data, we will require that prospective authors obtain permission
from the other teams before using the KMTNet data for publications.

\section{Future Improvements}
\label{sec:improve}

Following the work reported here creating and implementing
the 2015 event-finder algorithm,  and based on this experience,
numerous improvements are already being implemented for the 2016,
2017 and 2018 versions of this algorithm. 
They will be described more fully in the 2016 (and subsequent
data release paper(s).

The first improvement is simultaneous fitting to data from all three
observatories.  This is straightforward in principle because
all three observatories have identical input star catalogs,
so one can easily cross-identify light curves of the same physical
``star'' (really ``catalog star'', which very frequently is actually
a multi-star asterism).

\begin{figure}
\plotone{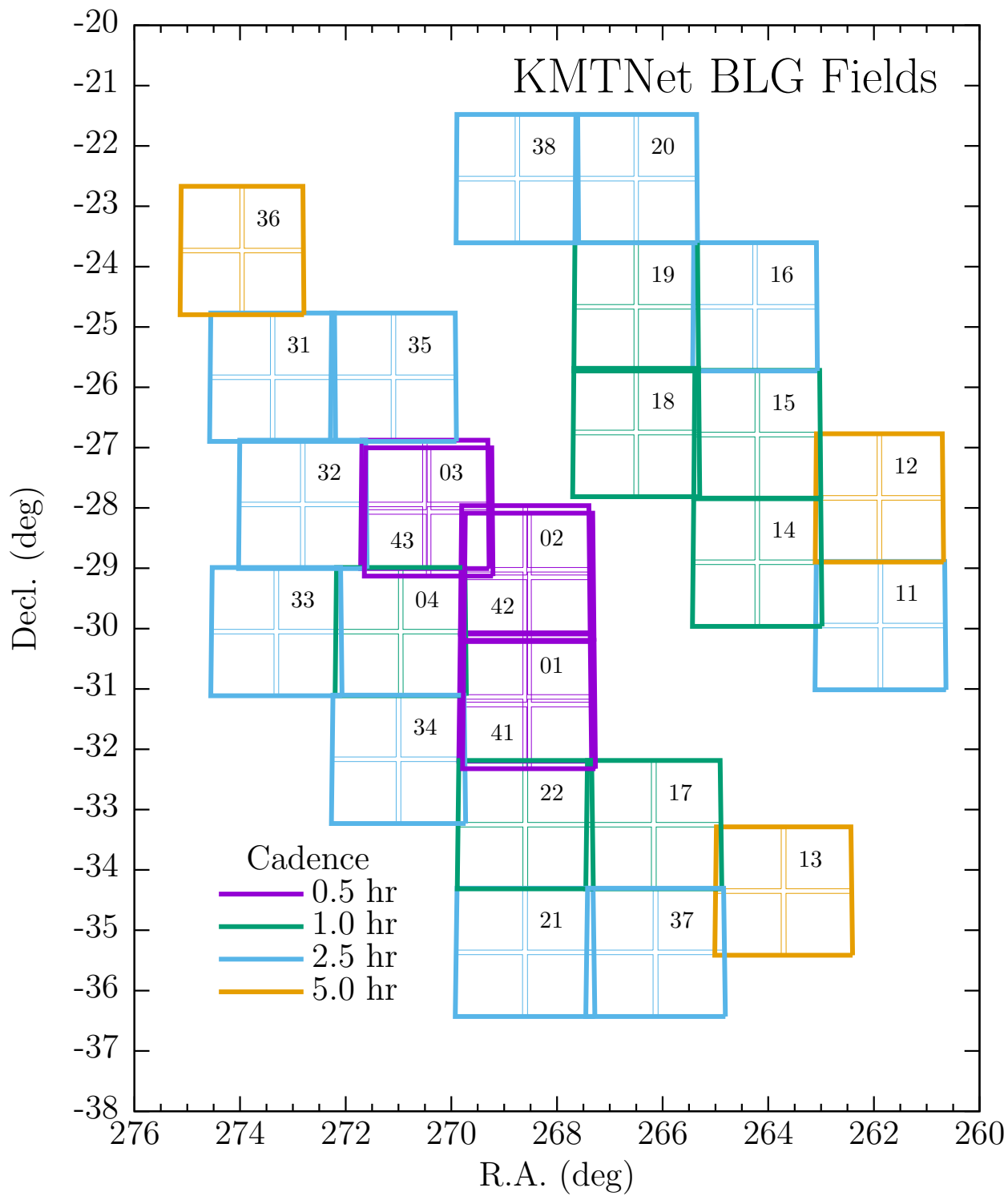}
\caption{27 fields observed by KMTNet in 2016, color-coded by cadence.  Note
that BLG(41,42,43) are shifted by $6^\prime$ relative to BLG(01,02,03),
which enables $\Gamma=2\,{\rm hr}^{-1}$ coverage of the chip gaps
while still preserving $\Gamma=4\,{\rm hr}^{-1}$ cadence over most
of this prime area.  Note also that a small area is covered by four fields
BLG(02,03,42,43) and so has $\Gamma=8\,{\rm hr}^{-1}$.
(Courtesy of Matthew Penny).
}
\label{fig:penny}
\end{figure}

The second is simultaneous fitting of overlapping fields.  This was
completely unnecessary in 2015 because there were no overlapping
fields.  However, in 2016, each of the three ``prime fields'' BLG01,
BLG02, BLG03, was also observed, slightly shifted, as BLG41, BLG42,
BLG43, in order to cover the gaps between chips.  Further, fields
BLG02 and BLG03 are now somewhat overlapping in order to obtain a very
high cadence on a small area $\Gamma=8\,{\rm hr}^{-1}$.  Hence, there
can be as many as four overlapping fields (see Figure~\ref{fig:penny}).
And this also means (taken together with the previous paragraph) that
there can be as many as 12 light curves that might be combined.  In
order to reliably cross-identify stars, however, this improvement will
only be implemented for OGLE-III catalog stars.

In 2015, we did not consider events with $t_\eff>99\,$days because
we had no out-of-year baseline data.  This will remain so in 2016
because the fields have been changed between 2015 and 2016, but also
because the artifacts in the 2015 data make them unsuitable as
a baseline of comparison.  Such longer events can only be
investigated beginning 2017.

On the other hand, it should be possible to extend the search
to much shorter timescales than the 2015 limit, $t_\eff\geq 0.56\,$days.
In 2015, such a search was not possible because of the large
number of short timescale artifacts.  It is likely that in 2016,
the search will be conducted independently in the $t_\eff\geq 1\,$day
and $t_\eff< 1\,$day domains, to guard against long timescale
events being ``killed'' by remaining short timescale artifacts.
Finally, we note that these short timescale searches should resolve
the problem found in 2015 searches of missing binary events characterized
primarily by very fast rising caustics.

Experience with the 2015 data shows that it is quite difficult to
reliably identify microlensing events based on the current threshold
$(\Delta\chi^2>500)$.  In 2015, we tentatively identified events from
KMTC data only based on this threshold, but then relied on data from
other observatories to confirm them, so the true $\Delta\chi^2$ was
significantly higher.  For 2016+, we are already using all available
data for 1) OGLE-III catalog stars in all fields and 2) all stars
outside of the six fields mentioned in the previous paragraph.  For
these, we will therefore demand $(\Delta\chi^2>1000)$.  For the
remainder (i.e., non-OGLE-III stars in BLG01, 02, 03, 41, 42, 43), we
will continue to demand $(\Delta\chi^2>500)$.

In addition, the process of visual inspection of 2015 light curve was
quite laborious. For 2016, we added
automated rejection of published variables, as well as variables
that were detected by KMTNet in 2015.  This will remove most
variables from the prime fields.  For the new fields (of which there
are 19), as well as the non-OGLE-III catalog stars in the prime
fields, we will have to flag the variables in 2016, but then will be
able to automatically censor them in future years. In 2017, we extended
the rejection of cataloged KMTNet variables to all fields and
added (after suitable testing) a rejection of previously found
``artifacts'' as well.  We also introduced a new algorithm
to eliminate a broad class of periodic variables,
which is not a periodogram, but whose details will be described
elsewhere.  And we introduced a new algorithm to remove short
timescale artifacts that masqueraded as short microlensing events.
Again, this will be described elsewhere.  Finally, for 2018,
we are introducing injection of fake microlensing events into
the process in order to measure the efficiency as a function
of event characteristics of the entire selection process, both
machine and human.  See for example, \citet{gould06}.

\section{{Summary}}
\label{sec:discuss}

We have presented a new algorithm for finding ``completed''
microlensing events and described its specific application to the
four, primary 2015 commissioning fields of the KMTNet microlensing
survey. We find 660 ``clear microlensing'' events and 182 ``possible
microlensing'' events discovered and assessed by our method. The light
curves of these events will be made publicly available on our website
(see Section~\ref{sec:website}) according to the data policy described
in Section~\ref{sec:policy2015}.  In Section~\ref{sec:policy}, we have
presented our general approach to KMTNet data releases, in addition to
the specific implementation of this approach for 2015 data.  Feedback
on the data products, data policy, etc., based on the use of these
data will be helpful in preparing future releases.

Our procedure for vetting microlensing candidates still relies heavily
on human inspection of the microlensing candidates. Thus, we also
discuss potential modifications for future years when we expect the
data to be substantially cleaner. In fact, work is currently underway
to improve this algorithm for the 2016 data. Finally, we note that
although the procedure for eliminating false positives will remain
dependent on the specific characteristics of the data, the algorithm
itself as presented in Section \ref{sec:algo} is completely general
and could be applied to diverse microlensing data sets.  Indeed,
\citet{ukirt16} have already applied this algorithm to find the first
infrared-only microlensing events from their UKIRT survey of heavily
extincted bulge regions.

\acknowledgments

We thank the OGLE collaboration for making available full information
about their 2015 event detections via their website
http://ogle.astrouw.edu.pl/, which enabled systematic checks of
our event-finder.
Work by YKJ, WZ and AG was 
supported by AST-1516842 from the US NSF. 
Work by IGS and AG was supported by
JPL grant 1500811.
This research has made use of the KMTNet system operated by the Korea
Astronomy and Space Science Institute (KASI) and the data were obtained at
three host sites of CTIO in Chile, SAAO in South Africa, and SSO in
Australia.

\end{document}